\documentclass{aa}
\usepackage{natbib,graphicx}
\newcommand{\Msun}{\mbox{\rm M$_{\odot}$}}
\begin{document}

\title{A synthetic view on structure and evolution of the Milky Way}

   \author{A.C. Robin
        \inst{1}
        \and
        C. Reyl\'e \inst{1}
	\and
	S. Derri\`ere \inst{2}
	\and
	S. Picaud \inst{1}
           }
   \institute{CNRS UMR6091, Observatoire de Besan{\c c}on, BP1615, 
    F-25010 Besan{\c c}on Cedex, France\\
		\email{annie.robin@obs-besancon.fr,celine@obs-besancon.fr,picaud@obs-besancon.fr}
	\and
	      CNRS UMR7550, Observatoire de Strasbourg, 11 rue de l'Universit\'e, 
    F-67000 Strasbourg, France\\
 \email{derriere@astro.u-strasbg.fr}}

        \offprints{Annie C. Robin}
   \date{Received 14 October 2002 / Accepted 11 July 2003}

   \titlerunning{Structure and evolution of the Milky Way}

   \abstract{Since the Hipparcos mission and recent large scale surveys 
in the optical and the near-infrared, new constraints have been 
obtained on the structure and evolution history of the Milky Way.
The population synthesis approach is a useful tool to interpret such
data sets and to test scenarios of evolution of the Galaxy. 
We present here new constraints
on evolution parameters obtained from the Besan\c{c}on model of population
synthesis and analysis of optical and near-infrared star counts. 
The Galactic potential is computed self-consistently, in agreement 
with Hipparcos
results and the observed rotation curve. Constraints are posed on the
outer bulge structure, the warped and flared disc, the thick disc and 
the spheroid populations. The model is tuned to
produce reliable predictions in the visible and the near-infrared in wide
photometric bands from U to K. Finally, we describe 
applications such as photometric and astrometric simulations and 
a new classification tool based on
a Bayesian probability estimator, which could be used in the framework 
of Virtual Observatories. As examples, samples
of simulated star counts at different wavelengths and directions 
are also given. 
   }

   \maketitle

\section{Introduction}

The population synthesis approach aims at
assembling current scenarios of galaxy formation and evolution, 
theories of stellar formation and evolution, models of stellar atmospheres and
dynamical constraints, in order to make a consistent picture
explaining currently available observations of different types (photometry, 
astrometry, spectroscopy) at different wavelengths.

The validity of any Galactic model is always questionable, as it describes 
a smooth Galaxy, while inhomogeneities exist, either in the disc
or the halo. The issue is not to make a perfect model that reproduces the
known Galaxy at any scale. Rather one aims to produce a useful tool to 
compute the 
probable stellar content of large data sets and therefore to test the 
usefulness of such data to answer a given question in relation to Galactic
structure and evolution. Modeling is also an effective way to test
alternative scenarios of galaxy formation and evolution.

The originality of the Besan\c{c}on model, as compared to a few other
population synthesis models presently available for the Galaxy, is the 
dynamical 
self-consistency. The Boltzmann equation allows the scale height
of an isothermal and relaxed population to be constrained by its velocity
dispersion and the 
Galactic potential. The use of this dynamical constraint avoids a set of 
free parameters quite difficult to determine: the scale height
of the thin disc at different ages. It gives the model an improved 
physical credibility.

In section 2, we show how the model describes the Galactic populations,
as they are accounted for in the model and how constraints
on galaxy evolution are incorporated.
In section 3 we describe current and future
applications of such a model { and give a sample of comparisons between model predictions and existing data}. In section 4 we discuss ongoing plans and 
future improvements. 

\section{Galactic populations and Galaxy evolution}

{ The scenario of evolution of the Galaxy slightly evolves as new 
data become available (new wavelengths, deeper samples, more accurate 
observations, ...), and as 
physical processes are better understood. Model
predictions have to be checked and eventually model parameters have to be
 refined. 
One should always take care that the model stays self-consistent and 
that the overall agreement is conserved. In this section
we describe the overall inputs in an up-to-date model of the Galaxy, 
how accurately the physical parameters are constrained, and which data are 
used to constrain them. This results in a description of the Galactic populations
consistent with the current scenario of Galaxy evolution.

The main scheme of the model is to reproduce the stellar content of the
Galaxy, using some physical assumptions and a scenario of formation
and evolution. We essentially assume that stars belong to four main 
populations : the thin disc (sect.2.1), the thick disc (sec.2.2), the 
stellar halo (or spheroid, sect. 2.3), and the outer bulge (sect. 2.4).
White dwarfs are taken into account separately but self-consistently 
(sect. 2.5).
A model of distribution of the extinction is also assumed (sect. 2.6).

The modeling of each population is based on a set of evolutionary tracks,
assumptions on density distributions, constrained either by dynamical
considerations or by empirical data, and guided by a scenario of
formation and evolution.}
 
In \cite{Bienayme1987a,Bienayme1987b} we have shown for the first time an
evolutionary model of the Galaxy where the dynamical self-consistency was
taken into account to constrain the disc scale height and the local
dark matter. Then 
\cite{Haywood1997} used improved evolutionary tracks and remote star
counts to constrain the initial mass function (IMF) and 
star formation rate (SFR) of
the disc population. The thick disc formation scenario has been studied
using photometric and astrometric star counts in many 
directions, which also provided its velocity ellipsoid, local density,
scale height, and mean metallicity. These physical constraints led to
a demonstration that the probable scenario
of formation was by an accretion event at early ages of the Galaxy \citep{Robin1996}.

More recently, the Hipparcos mission and large scale surveys in the optical
and the near-infrared have led to new physical constraints improving our
knowledge of the overall structure and evolution of the Galaxy. These
new constraints, now included in the present version
of the model, are presented below.

\subsection{Thin disc}
In the scheme of computing the population synthesis model, the stellar
content at each epoch has been computed from standard parameters such as
the initial mass function (IMF), a star
formation rate (SFR), and a set of evolutionary tracks \citep[see][and references 
therein]{Haywood1997}. 
The disc population has been assumed to evolve during 10 Gyr, age derived from
the white dwarf luminosity function { with an accuracy of about 15\% 
\citep{Wood1998}}.
Sets of IMF slopes and SFRs have been tentatively assumed and tested against
star counts. The preliminary tuning of the disc evolution parameters 
against relevant 
observational data was extensively described in \cite{Haywood1997}. We 
present below further {improvements in the disc description from 
new observational constraints}.

\subsubsection{Hipparcos inputs and the dynamical self-consistency}

Thanks to the Hipparcos mission, new and more accurate measurements 
have been made of the local density and 
potential \citep{Creze1998}, luminosity function \citep{Jahreiss1997}
and kinematics \citep{Gomez1997}. These new constraints have been
included in the modeling as follows. 

First, the new Hipparcos luminosity function is used to
constrain the IMF { at low masses} in the solar neighbourhood. The IMF is modeled
by a two-slope power law { (table~1)}. { At high masses, the slope $\alpha=3$ is slightly higher than the Salpeter assumption, as determined by \cite{Haywood1997}}. At the low mass end the slope
has been revised to $\alpha$=1.6 (instead of 1.7 in \cite{Haywood1997}) to 
give a better fit to the 
Jahrei\ss  new luminosity function from the revised Catalogue of Nearby 
Stars (2001, private communication). { The new 
IMF slope gives a better fit to the local luminosity function 
in the range $6~<~M_V <~13$ as shown in figure~1. At the lower luminosity 
end, the local sample is still too poor to ascertain the IMF slope. This slope
is in good agreement with independent and detailed analysis from 
\cite{Kroupa2000} which is in favor of a slope  $\alpha=1.3 \pm 0.5$ for the 
mass range 0.08 to 0.5 \Msun. Kroupa adopted four different
slopes in four mass ranges. Contrary to us, in the mass range 0.5 to 
1.0 \Msun, Kroupa 
adopts $\alpha=2.3 \pm 0.3$, that is a Salpeter IMF. We estimate that 
presently available field
data do not give enough constraints to adjust an IMF with four different slopes
depending on the mass range. We have rather chosen to rely
on a simpler modeling (two slopes) of the IMF for the thin disc population, 
until complementary constraints are available.}

\begin{figure*}
\begin{center}
\begingroup%
  \makeatletter%
  \newcommand{\GNUPLOTspecial}{%
    \@sanitize\catcode`\%=14\relax\special}%
  \setlength{\unitlength}{0.1bp}%
\begin{picture}(3600,2160)(0,0)%
\special{psfile=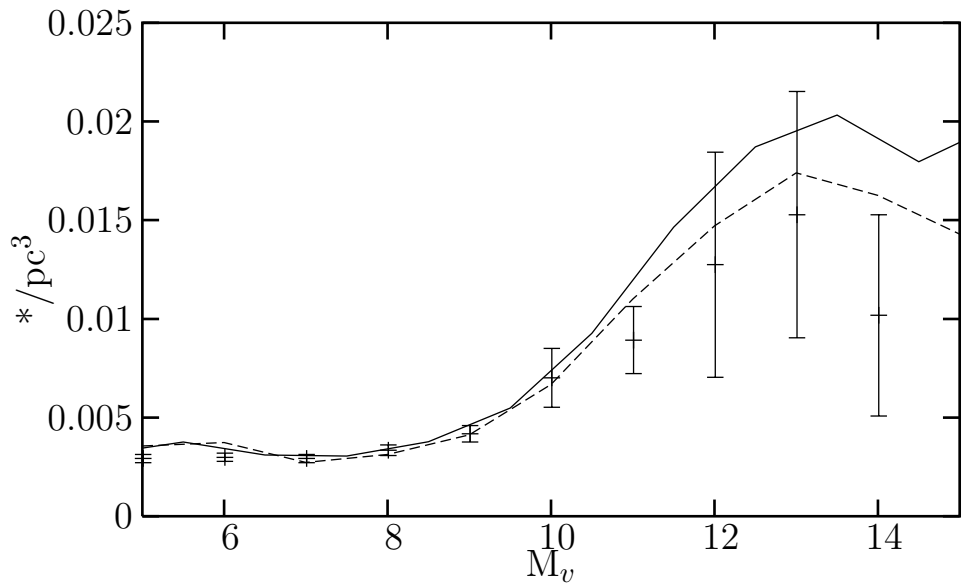 llx=0 lly=0 urx=360 ury=216 rwi=3600}
\fontsize{15}{\baselineskip}\selectfont
\put(1948,38){\makebox(0,0){M$_v$}}%
\put(74,1118){%
\special{ps: gsave currentpoint currentpoint translate
270 rotate neg exch neg exch translate}%
\makebox(0,0)[b]{\shortstack{*/pc$^3$}}%
\special{ps: currentpoint grestore moveto}%
}%
\put(3180,150){\makebox(0,0){ 14}}%
\put(2564,150){\makebox(0,0){ 12}}%
\put(1948,150){\makebox(0,0){ 10}}%
\put(1332,150){\makebox(0,0){ 8}}%
\put(716,150){\makebox(0,0){ 6}}%
\put(371,2086){\makebox(0,0)[r]{ 0.025}}%
\put(371,1714){\makebox(0,0)[r]{ 0.02}}%
\put(371,1342){\makebox(0,0)[r]{ 0.015}}%
\put(371,969){\makebox(0,0)[r]{ 0.01}}%
\put(371,597){\makebox(0,0)[r]{ 0.005}}%
\put(371,225){\makebox(0,0)[r]{ 0}}%
\end{picture}%
\endgroup
 
\caption{Luminosity function in the V band 
in the solar neighbourhood. Crosses with 
Poisson errors: Jahreiss determination from the revised Catalogue of Nearby
Stars; Lines: model LF assuming an IMF slope of $\alpha$=1.7 (solid line) and
1.6 (dashed line) at m$<$1\Msun.}
\end{center}
\end{figure*}

Second, as Hipparcos has provided new constraints on the local kinematics, 
the age-velocity
dispersion relation from \cite{Gomez1997} is introduced { (cf sect. 2.1.4)}.

 Third, the local density of interstellar matter has been revised. From 
\cite{Dame93}
the local surface density of HI+H$_2$ is about 6 $\Msun$ pc$^{-2}$ with a
considerable uncertainty of about a factor of 2. With a scale height estimated
to 140 pc, the local mass density is 0.021 $\Msun$ pc$^{-3}$. This is 
about half
the local density used in the previous version of our model. We 
notice however that the uncertainty is still very large.

{ Adding up the mass density from stars, interstellar matter and local 
contribution of the dark halo (see table~2), we end up with a} total local
mass density of 0.0759  $\Msun$ pc$^{-3}$, in agreement with the 
dynamical value 0.076 $\pm 0.015 \Msun \mathrm{pc}^{-3}$ 
obtained by \cite{Creze1998} { from a detailed analysis of the K$_z$ force 
on a sample of A and F stars observed by Hipparcos}.

\begin{table*}
\label{evolution}
{\centering
\caption{Age, metallicity ($[\frac{Fe}{H}]$) (mean and dispersion about the mean), radial metallicity gradient (dex/kpc), initial mass 
function (IMF), and star formation rate (SFR) of the stellar components.}
\begin{tabular}{|l|l|r|c|c|c|c|}
\hline
	&Age (Gyr)	&$[\frac{Fe}{H}]$ (dex)	&$\frac{d[Fe/H]}{dR}$	&IMF				&SFR\\
\hline
Disc	&0-0.15	  &0.01 $\pm$ 0.12		&& 				&\\
	&0.15-1	  &0.03 $\pm$ 0.12		&	&				&\\
	&1-2	  &0.03 $\pm$ 0.10		&	&dn/dm $\propto$ m$^{-\alpha}$	&\\
	&2-3	  &0.01 $\pm$ 0.11		&$-$0.07	&\hspace{0.3cm}$\alpha$ = 1.6 for m $< 1 \Msun$&constant\\
	&3-5	  &$-$0.07 $\pm$ 0.18		&	&\hspace{0.3cm}$\alpha$ = 3.0 for m $> 1 \Msun$	&\\
	&5-7	  &$-$0.14 $\pm$ 0.17		&	&				&\\
	&7-10	  &$-$0.37 $\pm$ 0.20		&	&				&\\
\hline
Thick disc	&11	&$-$0.78 $\pm$ 0.30	&0.00	&dn/dm $\propto$ m$^{-0.5}$	&one burst\\
\hline
Stellar halo	&14	&$-$1.78 $\pm$ 0.50	&0.00	&dn/dm $\propto$ m$^{-0.5}$	&one burst\\
\hline
Bulge		&10	&0.00 $\pm$ 0.40	&0.00	&dn/dm $\propto$ m$^{-2.35}$ for m $>0.7 \Msun$	&one burst\\

\hline
\end{tabular}\par}
\end{table*}

\begin{table}[h]
{\centering
\caption{Local mass density $\rho_0$ of the stellar components, the dark matter
halo and the interstellar medium (ISM). $W$-velocity dispersion $\sigma_W$ used for
the dynamical self-consistency, and
disc axis ratios $\epsilon$ resulting from this process.
The white dwarf mass density is computed assuming a white dwarf mass of 
0.6$\Msun$.}
\label{local_density}
\begin{tabular}{lllll}
\hline
	&Age	&$\rho_0$		&$\sigma_W$	&$\epsilon$\\
	&(Gyr)	&($\Msun \mathrm{pc}^{-3}$)	&(km s$^{-1})$	&\\
\hline
Disc	&0-0.15	&4.0$\times$10$^{-3}$	&6		&0.0140\\
	&0.15-1	&7.9$\times$10$^{-3}$	&8		&0.0268\\
	&1-2	&6.2$\times$10$^{-3}$	&10		&0.0375\\
	&2-3	&4.0$\times$10$^{-3}$	&13.2		&0.0551\\
	&3-5	&5.8$\times$10$^{-3}$	&15.8		&0.0696\\
	&5-7	&4.9$\times$10$^{-3}$	&17.4		&0.0785\\
	&7-10	&6.6$\times$10$^{-3}$	&17.5		&0.0791\\
  &WD & 3.96$\times$10$^{-3}$	\\
\\
Thick disc & 11	&1.34$\times$10$^{-3}$	&		&\\
   &WD & 3.04$\times$10$^{-4}$ \\
\\
Stellar halo   & 14	&9.32$\times$10$^{-6}$ &	&0.76	\\
\\
Dark matter halo & &9.9$\times$10$^{-3}$&		&1.\\
ISM	&	&2.1$\times$10$^{-2}$	&		&\\
\hline
\end{tabular}\par}
\end{table}

The evolution model gives the present distribution of stars in a column 
of unit volume centered on the sun as a function of intrinsic parameters 
(age, mass, effective temperature, gravity, metallicity). Since the 
evolution model does not yet account for orbit evolution, we 
redistribute the stars in the reference volume over the $z$ axis according 
to the age/thickness relation deduced from the Boltzmann equation.

\begin{table*}
\label{density}
{\centering
\caption{Density laws and associated parameter of the stellar components.
$a^2 = R^2+\frac{z^2}{\epsilon^2}$ where $R$ is the galactocentric distance, 
$z$ is the height above the Galactic plane, and $\epsilon$ is
the axis ratio. Values of $\epsilon$ are given 
in table~2. The disc density law is given here without the warp and
flare. 
The corrections for these structures are given in the text
(sect. 2.1.3). 
{ d$_0$ are normalization constants. For the bulge, x,y and z are in
  the bulge reference frame and values of N, ${x_0}, y_0,
  z_0$ 
and $R_c$ are given in the text (sect. 2.4).}}
\begin{tabular}{lllll}
\hline
	   &\multicolumn{1}{c}{density law} &\\
\hline
Disc	   &$\rho_0/d_0 \times \{\exp(-(a/h_{R_+})^2)-\exp(-(a/h_{R_-})^2)\}$ &if age $\leq$ 0.15 Gyr\\
 &\hspace{0.3cm}with $h_{R_+}$ = 5000 pc, $h_{R_-}$ = 3000 pc & \\
&$\rho_0/d_0 \times \{\exp(-(0.5^2+a^2/h_{R_+}^2)^{1/2})-\exp(-(0.5^2+a^2/h_{R_-}^2)^{1/2})\}$ &if age $>$ 0.15 Gyr\\ 
&\hspace{0.3cm}with $h_{R_+}$ = 2530 pc, $h_{R_-}$ = 1320 pc\\
Thick disc &$\rho_0/d_0 \times \exp{(-\frac{R-R_\odot}{h_{R}})} \times (1-\frac{1/h_{z}}{x_l\times (2.+x_l/h_{z})}\times z^{2})$     &if $|z|\leq x_l$, $x_l$=400 pc\\        
&$\rho_0 \times \exp{(-\frac{R-R_\odot}{h_{R}})} \times 
\frac{\exp(x_l/h_z)}{1+x_l/2h_z}\exp({-\frac{|z|}{h_{z}})}$        &if $|z|>x_l$\\
&\hspace{0.3cm}with $h_R$ = 2500 pc, $h_z$ = 800 pc &\\
Spheroid  &$\rho_0/d_0 \times (\frac{a_c}{R_\odot})^{-2.44}$      &if $a \leq a_c$, $a_c$ = 500 pc\\
  &$\rho_0 \times (\frac{a}{R_\odot})^{-2.44}$    &if $a > a_c$\\
Bulge &$N\times \exp (-0.5\times r_s^2)$ & $\sqrt{x^2+y^2} < R_c$ \\
 & $N\times \exp (-0.5\times r_s^2)\times \exp (-0.5 (\frac{\sqrt{x^2+y^2}-R_c}{0.5})^2)$ & $\sqrt{x^2+y^2} > R_c$\\
 & \hspace{0.3cm}with
$r_s^2=\sqrt{[(\frac{x}{x_0})^2+(\frac{y}{y_0})^2]^2+(\frac{z}{z_0})^4}$
\hspace{0.3cm} \\
ISM &$\rho_0 \times exp(-\frac{R-R_\odot}{h_{R}}) \times exp(-\frac{|z|}{h_{z}})$	& \\
&\hspace{0.3cm}with $h_{R}$ = 4500 pc, $h_{z}$ = 140 pc \\
Dark halo & $\frac{\rho_c}{(1.+(a/R_c)^{2})}$ with $R_c$ = 2697 pc and $\rho_c$ = 0.1079\\

\hline
\end{tabular}\par}
\end{table*}

With these new inputs, the dynamical self-consistency process is applied as
first described in \cite{Bienayme1987a}: assuming that the Galactic
potential is given by the sum of the stellar populations (those in the present
model), and adding up the interstellar matter and the dark matter halo, 
we compute
the potential through the Poisson equation. The disc is not modeled with a 
double
exponential but with Einasto laws \citep{Einasto}, ensuring continuity and derivability in the Galactic plane. The formulae are given in table~3. 
The thin disc population is sliced
into seven isothermal populations
of different ages, from 0 to 10 Gyr. Each sub-population (except the youngest
one, which cannot be considered relaxed) has its velocity
dispersion imposed by the age/velocity dispersion relation. Then using
equation~\ref{bol} we compute its scale height (or axis ratio in case of ellipsoids),  
while the rotation curve constrains the shape and local density of the dark
matter halo \citep{Bienayme1987a}. 
\begin{center}
\begin{equation}
\label{bol}
\frac{\rho(z)}{\rho(0)} = \mbox{exp}\left(-\frac{\Delta\Phi}{\sigma_{W}^2}\right)
\end{equation} 
\end{center}

A new total mass density is then computed. The process is iterated until the 
values of the potential and the disc thicknesses change by less than 1\%.
The resulting disc axis ratios $\epsilon$ are given in table~2. The rotation
curve of the best fit model is shown in figure~2.

{ The local density of the dark matter halo amounts to $9.9~\times~10^{-3}$~\Msun pc$^{-3}$, about 20\% higher than the value determined by \cite{Bienayme1987a}, while the mass density of our disc
is lower. The difference comes from the fact that they included an ``unseen mass disc'' of local density $0.01~$\Msun pc$^{-3}$ of which existence has been definitely ruled out by \cite{Creze1998}. The dark halo serves
to compensate for the lack of force coming from the disc, in order to adjust the rotation curve. This is the major massive component at galactocentric
distances larger than 5 kpc. The question of the composition of this dark
component is still open, probably mainly non baryonic, but not exclusively. 
We shall consider in sect. 2.5 the possibility of this component to be
made of ancient white dwarfs.}

{ The overall model mass density ends up with the following values: the outer 
bulge mass, coming from old populations in the central part of the Galaxy, 
is $2.03\times~10^{10}$ \Msun, excluding the central black hole and clusters. 
The mass
of the interstellar matter amounts to 4.95$\times~10^{9}$ \Msun in our modeling.
The stellar thin disc reaches a mass of 2.15$\times~10^{10}$ \Msun, 
the thick disc 3.91$\times~10^{9}$ \Msun and the
stellar halo 2.64$\times~10^{8}$ \Msun (excluding eventual ancient halo white 
dwarfs which belong to the ``dark halo''). The dark halo has a mass of 
4.53$\times~10^{11}$ \Msun inside the sphere of 50 kpc radius. 
The overall Galaxy mass to R$<$50 kpc amounts to 5.04$\times~10^{11}$ \Msun,
and 9.97$\times~10^{11}$ \Msun within 100 kpc.
These values are in good agreement with 
the detailed analysis of the Galactic mass by \cite{Sakamoto2003},
from the latest kinematics
of Galactic satellites, halo globular clusters and horizontal
branch stars, which estimates the Galactic mass to $5.5^{+0.1}_{-0.4} 
\times 10^{11}$ \Msun within 50 kpc and about $10\times 10^{11}$ \Msun within 
100 kpc.}

\begin{figure}
\begin{center}
\includegraphics[width=7cm,clip=,angle=-90]{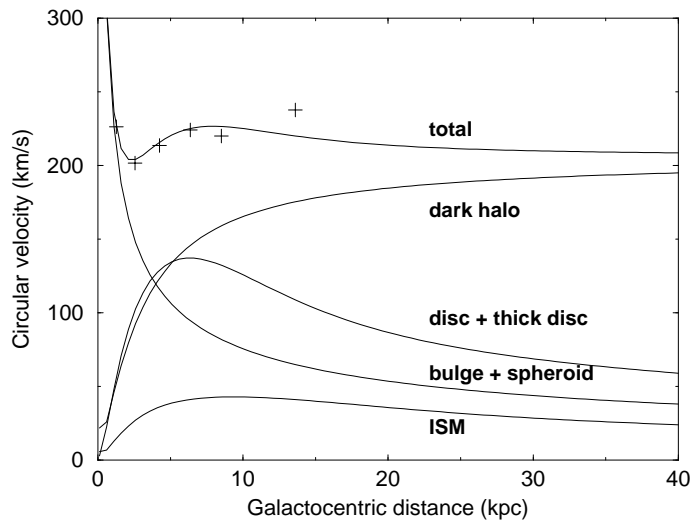}
\caption{Circular velocity of the self-consistent model. Crosses are 
observational constraints from  \protect\cite{Caldwell1981}.  
Contributions from main Galactic components are also given.}
\end{center}
\end{figure}

The density laws resulting from self-consistent adjustments of the Galactic 
potential are finally used to compute the
stellar content all over the Galaxy, assuming that the disc IMF and SFR
at the solar neighbourhood are suitable for the entire disc population.
{ The inferred dark matter local density is also used to compute the 
contribution of potential ancient white dwarfs in star counts (see 
sect. 2.5.3.)}

\subsubsection{Atmosphere models and colour estimates}

The evolutionary model computes the distribution of stars in the space 
of intrinsic parameters: effective temperature, gravity, absolute
magnitude, mass and age. { This distribution transcribes the theoretical
constraints coming from stellar and galactic evolution (mainly the IMF, the SFR,
the evolutionary tracks, and the metallicity distributions) into 
observable statistics, with the use of suitable atmosphere models.} 

The metallicity is estimated from the age of each component, assuming an
age/metallicity relation and allowing for a Gaussian dispersion about the
mean. For the thin disc, we use \cite{Twarog80} 
age/metallicity distribution (mean and dispersion about the mean, see 
table~1). 
An alternative new metallicity distribution from \cite{Haywood2001}, more
centered on the solar metallicity, can also be used. However important to
understand the scheme of the Galaxy chemical evolution, the differences between Twarog and Haywood
distributions have only marginal impact on star count predictions as
far as broad photometric bands are used. { Conversely, it is not possible
to settle this issue with ordinary photometric star counts.}

The theoretical parameters are converted into
colours in various photometric systems through stellar atmosphere models 
corrected to fit empirical data 
\citep{Lejeune1997,Lejeune1998} { (the BaSeL~2.2 data base)}. While some uncertainties
still remain in the resulting colours for extreme spectral types, { the BaSeL data sets allow an efficient and consistent conversion of theoretical parameters
(effective temperature, gravity and metallicity) to colors in various photometric systems}. For low mass stars,
synthetic colours from \cite{Baraffe1998} have been used in the near-infrared
in place of Lejeune calibrations. 
Empirical densities, infrared luminosities and colours of giants of 
spectral types M7 and later are introduced following
the observations of \cite{These_Guglielmo}. The mid-infrared modeling of these
late M giants and AGB stars may be inaccurate. It will be revised in the
near future.

At present, calibrations are available
for the Johnson-Cousins UBVRIJHKL. The GAIA G magnitude can also be simulated.
The Sloan SDSS { and Str\"omgren photometric systems are} planned.

\subsubsection{External disc}

Several studies have shown that the edge of the disc is detected at a 
galactocentric 
distance of about 14 kpc \citep{Robin1992,Ruphy1996}.
However when analysing 
DENIS near-infrared survey data \citep{Epchtein} in the Galactic plane, 
we have shown that the warp and 
flare of the external disc appear clearly in near-infrared star counts
\citep{Derriere2001a}. {For a long time the warp and flares have been clearly 
seen in HI data \citep{Henderson1982, Burton1986, Burton1988, Diplas1991}, in 
molecular clouds \citep{Grabelsky1987,Wouterlout1990,May1997}, from OB stars 
\citep{Miyamoto1988,Reed1996} and finally in COBE data 
\citep{Porcel1997,Freudenreich1998,Drimmel2001}. The existence of such a warp in old stellar
populations puts constraints on the time scale of this structure. However there
is evidence that the stars do not follow exactly the same warp
as the gas \citep{Djorgovski1989,Porcel1997}. Possible origins of the warps are
still under discussion but may be resolved by suitable N-body simulations 
\citep{Binney1992}. Warps may have originated from interactions between the disc 
and (i) the dark halo (if angular momenta are not aligned), (ii) nearby 
satellite galaxies, such as the Sagittarius dwarf or the Magellanic clouds, 
(iii) infalling intergalactic gas. From the analysis of their angular momenta,
 \cite{Bailin2003} argues that the Milky Way warp and the Sagittarius Dwarf 
Galaxy may be coupled. \cite{Garcia-Ruiz2002} have studied
by N-body simulations the possibility for the warp to be caused by the tides of the Magellanic Clouds. They find that neither orientation nor amplitude
of the warp can be reproduced in this way. Alternatively \cite{López-Corredoira2002} propose that the warp caused by intergalactic accretion flows onto the Milky Way disc.

Despite the uncertainty on the warp's origin, in the modeling process we have started with warp and flare parameter values close to the ones observed in the gas and molecular clouds. Then, simulations have been compared with
near-infrared data from the DENIS survey in a set of relevant directions, and
warp and flare parameters have been adjusted. The parameters are 
described below :}

For the warp, a tilted ring model such
as \cite{Porcel1997} is considered. When computing
density laws for a warped disc, the galactocentric
coordinates ($R$,$\theta$,$z$) 
at galactocentric radii larger that $R_\mathrm{warp}$,
are shifted
perpendicular to the plane by a value 
$$z_\mathrm{warp} = z_c \cos (\theta -\theta_\mathrm{max})$$
 with 
$\theta_\mathrm{max}$ and $z_c$ being respectively the direction
in which the warp is maximum, and the maximum elevation of the 
warp at a given radius. We adopt a linear increase of the
warp amplitude $z_c$, such as 
$$z_c = \gamma_\mathrm{warp} (R-R_\mathrm{warp})$$

Following \citet{Burton1988}, we assume
that the Sun lies approximately on the line of nodes of the warp 
($\theta_\mathrm{max} = 90^\circ$). 
We adopt an amplitude similar to \citet{Gyuk99} 
($\gamma_\mathrm{warp}=0.18$). Preliminary results obtained from the
analysis of a part of the DENIS survey
\citep{Derriere2001a,Derriere2001b} indicate a best
value for the starting galactocentric radius of the warp to be 
$R_\mathrm{warp}=8.4$~kpc, close to $R_\odot$.

As in \citet{Gyuk99}, we model the flaring of the disc
by increasing the scale height by a factor $k_\mathrm{flare}$,
beyond a galactocentric radius $R_\mathrm{flare}$, with
$$
    k_\mathrm{flare} = 1 + \gamma_\mathrm{flare}(R-R_\mathrm{flare})
$$
  
The surface density in
the flaring disc is the same as in the normal one, leading to correction
factors in the density laws. 

The amplitude is taken from \citet{Gyuk99} 
($\gamma_\mathrm{flare}=5.4 \times 10^{-4}$ kpc$^{-1}$).
The radius at which the disc starts flaring lies beyond the solar
circle. { However \cite{Derriere2001b} found that the minimum radius of the flare might depend on the longitude considered. He determined a preliminary average value of
$R_\mathrm{flare}=9.5$~kpc which may be reconsidered when near-infrared surveys such as DENIS and 2MASS will be completely analysed. }

An example of the warp and flare effects on star counts in the near 
infrared is shown in
figure~3. { Squares are DENIS K star counts on a strip crossing 
the plane at 
l=266.1$^o$ and dashed lines are model predictions in same directions. 
The peak of the model predictions without
warp and flare (top panel) is displaced from the data peak.
The middle panel shows predictions including the warp but no flare. The 
Galactic warped plane is well
placed but the peak is too sharp. In the bottom panel the model including
both features (warp and flare) correctly reproduces the star counts.}

\begin{figure*}  
\centering
\label{preston}
\includegraphics[width=12cm,clip=,angle=0]{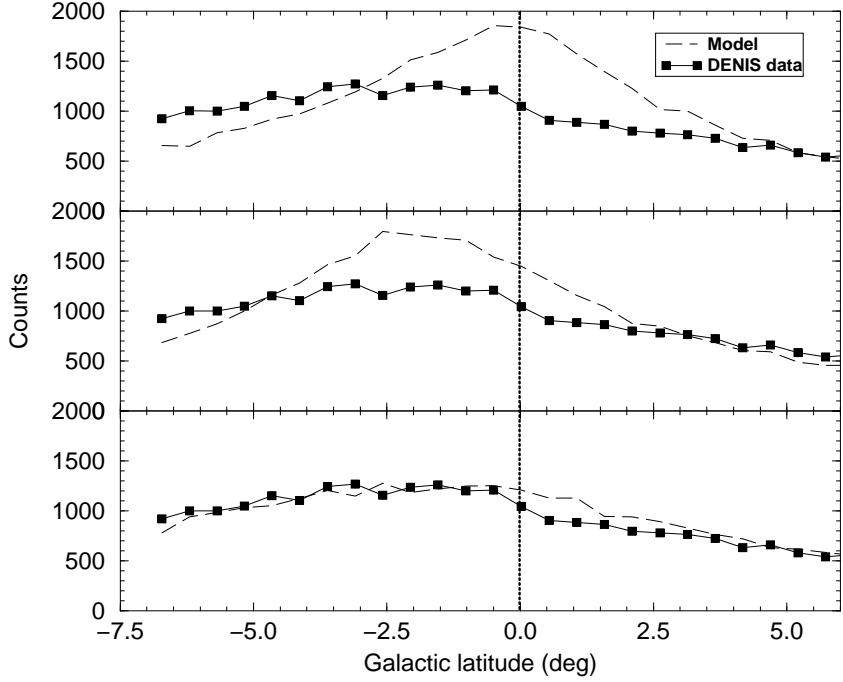}
\caption{Star counts in K along a DENIS strip in counts per
  0.8$^o$~bin in 
declination (this strip crosses the plane at l=266.1$^o$). Data
are in solid lines, models in dashed lines. Top panel: flat
disc. Middle 
panel: warped disc. Bottom panel: warped and flaring disc.}
 \end{figure*}

\subsubsection{Disc kinematics}

The disc velocity ellipsoid is taken from \cite{Gomez1997}, deduced
from Hipparcos data analysis. Velocity dispersions as a function of age and 
galactocentric gradients are given in Table~4.
The asymmetric drift is computed following the relation :

$V_{\scriptsize\mbox{ad}} = \frac{\sigma_U^2}{2 V_{\scriptsize\mbox{LSR}}}
\{\frac{R}{\rho} \frac{d\rho}{dR} + 2\times \frac{R}{\sigma_{U}} \frac{d\sigma_{U}}{dR} + (1 - 
\frac{\sigma_V^2}{\sigma_U^2}) + (1 - \frac{\sigma_W^2}{\sigma_U^2})\}$

where $\frac{d\rho}{dR}$ is the density gradient.
The Sun velocities are $U_\odot$ = 10.3 km s$^{-1}$, $V_\odot$ = 6.3 
km s$^{-1}$, $W_\odot$ = 5.9 km s$^{-1}$, and $V_{\scriptsize\mbox{LSR}}$ = 
226 km s$^{-1}$. The later value is obtained after fitting the mass 
model to the observed rotation curve (see sect. 2.1.1).
Asymmetric drift values computed at the solar position are given in 
table~4.

\begin{table*}
\label{cinematique}
{\centering
\caption{Velocity dispersions ($\sigma_U,\sigma_V,\sigma_W$), asymmetric drift 
$V_{\scriptsize\mbox{ad}}$ at the solar position (see section 2.1.4)
  and 
velocity dispersion gradient $\frac{d\ln(\sigma_{U}^2)}{dR}$, where
$\sigma_{U}$ is expressed in km $s^{-1}$ and R in kpc.}
\begin{tabular}{|l|l|c|c|c|c|c|}
\hline
	&Age	&$\sigma_U$	&$\sigma_V$	&$\sigma_W$	
&$V_{\scriptsize\mbox{ad}}$	&$\frac{d\ln(\sigma_U^2)}{dR}$	
\\
	&(Gyr)	&(km s$^{-1})$	&(km s$^{-1})$	&(km s$^{-1})$  
&(km s$^{-1})$	&	\\
\hline
Disc	&0-0.15	&16.7	&10.8		&6	& 3.5 & \\
	&0.15-1	&19.8	&12.8		&8	& 3.1	&\\
	&1-2	&27.2	&17.6		&10	& 5.8	&\\
	&2-3	&30.2	&19.5		&13.2	& 7.3	 &	$-2 \times 10^{-1}$	\\
	&3-5	&36.7	&23.7		&15.8	& 10.8	&\\
	&5-7	&43.1	&27.8		&17.4	& 14.8	&\\
	&7-10	&43.1	&27.8		&17.5	& 14.8	&\\
\hline
Thick disc &	&67	&51		&42	&53	&0\\
\hline
Spheroid   &	&131	&106		&85	&226	&0\\
\hline
Bulge	   &	&113	&115		&100	&79	&0\\
\hline
\end{tabular}\par}
\end{table*}

\subsection{Thick disc}

The thick disc formation scenario that drives the model characteristics
is a formation
by a merger event (or several merger events) at the beginning of the
life of the thin disc. This scenario well explains the observed properties of
this population : The accretion body heats the stellar population 
previously formed in the disc making a thicker 
population with larger velocity dispersions. Due to the epoch of this
event, the thick disc abundances are intermediate between the stellar halo and
the present thin disc. 
{ Chemical abundance measurements
imply that the duration of star formation in the thick disc cannot 
be larger than 1 Gyr \citep{Mashonkina2001}.
The estimated epoch of thick disc formation in the literature ranges
between 8 Gyr \citep{Ibukiyama} and
13 Gyr \citep{Pettinger} from chemical considerations, while \cite{Quillen2001}
find 9 Gyr from kinematics. We have finally adopted an age of 11 Gyr for 
this population. This is just slightly older than the thin disc, and younger
than the stellar halo. The
choice of the slightly different age (by 1 Gyr) would not change 
significantly wide band photometric star counts.
Other observational constraints come from kinematics : the velocity ellipsoid
and asymmetric drift have been measured by radial velocity or proper motion
distributions. 
\cite{Ojha1996,Ojha1999} determined the velocity ellipsoid of the
thick disc from an analysis of a photo-astrometric survey 
in three directions at intermediate latitudes (center, anticentre,
and antirotation). They found the velocity dispersions to be 
($\sigma_U, \sigma_V, \sigma_W$) = (67, 51, 42) in km s$^{-1}$ with an 
asymmetric drift of 53 km s$^{-1}$. These values are in close agreement 
with a recent determination
by \cite{Soubiran2003} who find a rotational lag of 51 km s$^{-1}$, and a
velocity ellipsoid ($63\pm 6, 39\pm 4, 39\pm 4$) km s$^{-1}$.
The asymmetric drift is not found to have any vertical gradient inside
this population. The eventual radial gradient is still undetermined and 
is assumed to be zero in modeling. The absence of chemical and kinematical
gradients and the large gap in properties between the thick disc and 
the stellar
halo (scale height, rotational velocity) rule out the hypothesis that the thick disc has been formed
during a dissipational collapse \citep{Majewski1993} and rather favour a formation
by a strong heating of a preliminary thin disc induced by a merging of a satellite onto the Galaxy \citep{Quinn1993,Robin1996}.

In order to model this population, a single epoch of
star formation (that is a period of formation negligible compared to the
age of the Galaxy) is assumed.} Oxygen enhanced evolutionary
tracks have been taken from \cite{Bergbush1992}. The horizontal
branch (HB) has been added following models from \cite{Dorman1992}.
Stars of effective temperature between 7410 K and 6450 K are assumed to 
be RRLyrae.
There is no blue horizontal branch, as seen on metal rich globular clusters. 
The position of the HB is assumed
to be 3.54 visual magnitudes brighter than the main sequence turnoff. 

The adopted thick disc metallicity is $-0.78$~dex,
in agreement with in situ spectroscopic determination from 
\cite{Gilmore1995} and photometric star count 
determinations \citep{Robin1996,Buser1999}. 
The low metallicity tail of the thick disc is not taken into account, but 
an internal metallicity dispersion of 0.3~dex is included. 
 
The thick disc density law has often been
assumed in the literature to be exponentially 
decreasing perpendicular to the Galactic plane. Tests of various laws have 
been performed by comparing data with model predictions in various directions.
At present time, it does not seem possible to obtain detailed constraints 
on the true shape, either exponential or sech$^2$, because the distances are
not known with sufficient accuracy. Moreover the contribution of short 
distance stars to the star counts is often negligible due to the small 
volume where they are counted. We have assumed that the
thick disc shape is a truncated exponential along the z axis: at 
large distances the law is exponential, at short distances it is a parabola 
(see table~3).  This option ensures the continuity 
and derivability of the density law in the plane, to ease the computation of
the thick disc contribution to the potential.

The density law parameters and IMF adopted are well constrained by 
the analysis of 
deep wide field and near-infrared star counts at high and intermediate 
Galactic latitudes \citep{Reyle2001a}.
The best fit model has a scale length $h_{R}$ = 2500 $\pm$ 500 pc, 
a scale height $h_{z}$ = 800 $\pm$ 50 pc with $x_l$=400 pc, and a local density 
$\rho_{0}$ = 1.03$\times 10^{-3}$ 
stars pc$^{-3}$ or 7.6$\times 10^{-4}$ $\Msun$ pc$^{-3}$
for M$_V$ $\leq$ 8, corresponding to 6.8\% of the thin disc local density.
The IMF is dN/dm $\propto$ m$^{-0.5}$, the slope being also constrained
by large data sets \citep{Reyle2001a}. The total density (excluding white
dwarfs) is then 2.83$\times 10^{-3}$ 
stars pc$^{-3}$ or 1.34 10$^{-3}$ $\Msun$ pc$^{-3}$.
This value does not account for binarity. Thus the true IMF should slightly
steepen. The binarity correction should play a role only in very deep star
counts. Such star counts are few and they remain
uncertain due to strong galaxy contamination.

\subsection{Stellar halo}

{ We here distinguish what we call the stellar halo, or spheroid, from the
dark matter halo. The spheroid is distinct from the bulge in the model
construction. However they may be related in their formation scenario. The
spheroid, or stellar halo, is essentially old and metal poor. The bulge is
also quite old but its abundance distribution is sensitively different, with 
a metallicity distribution
closer to the one of the thick disc population. 
On the other hand, the stellar halo should be considered distinct from 
the dark matter halo, which is probably mainly non baryonic, although 
a small part of it can be formed
of stellar remnants. We shall consider this point in sect. 2.5.3.}

Mainly two possible scenarios of formation of the spheroid are competing. 
The first
one is a dissipational collapse of pre-galactic gas, the second one being
based on accretion of smaller pieces. Streams have already been
found, related to known accretion events (Sagittarius galaxy, ...) but the
proportion of spheroid stars which can originate from such streams is still
not known. Therefore
our standard model assumes a homogeneous population of spheroid stars with a
short period of star formation, which should be a good representation
on large scales. Comparisons with true data should
then, by contrast, help finding existing streams. \cite{Bergbush1992}
oxygen enhanced models are used assuming a 
Gaussian metallicity distribution of mean  -1.78 and dispersion 0.5 dex.
The age of the spheroid is assumed to be 14 Gyr{, a value which may be
slightly high, as constrained by the
analysis of the WMAP experiment \citep{Bennet2003}. However, a change by 
0.5 Gyr would have no effects on star count predictions.}

The density law and IMF of the spheroid were constrained using deep star
counts at high and medium Galactic latitudes \citep{Robin2000}. 
This global adjustment of the spheroid shape has led to the following values:
the density
law (see table~3) is characterized by a power law index $n$ = 2.44, 
a flattening $\epsilon$ = 
0.76 and a local density $\rho_0$ = 2.185$\times 10^{-5}$ pc$^{-3}$ or 
0.932 10$^{-5}$ $\Msun$ pc$^{-3}$ including all red dwarfs down to 
the hydrogen burning limit, but not the white dwarfs. This local density 
has been
computed with an IMF dN/dm $\propto$ m$^{-0.5}$, similar to the globular 
clusters and
to the thick disc population (see discussion below). The total local density 
depends a lot on the
assumed IMF at low masses. With an IMF of dN/dm $\propto$ m$^{-2}$, the 
total local density would be\footnote{We 
apologize for a mistake in the
published paper \citep{Robin2000} where the given local densities of the
spheroid were wrong by a factor of 2}
$8.19 \times 10^{-5}$ stars~pc$^{-3}$. Model predictions are affected 
by the IMF slope value only
at magnitudes fainter than V$=22$ at the Galactic pole (see figure~4 and sect. 4).

The relative density of the spheroid to the disc is about 0.06\% when
one selects stars at M$_V<8$\footnote{the ratio computed with low mass stars
included would be completely dominated by the IMF slope used, parameter
which is not strongly constrained}. Yet this value ignores possible old 
white dwarfs.
The ancient white dwarf contribution will be considered separately below.

\subsection{Outer bulge and disc hole}

{ Bulge formation in the Galaxy is still an open question. Several scenarios compete
to explain the observed structures. The proposed scenarios are : primordial 
collapse \citep{Eggen1962}, hierarchical galaxy formation 
\citep{Kauffmann1994} (early merging), infall of satellite galaxies (late 
merging), and secular evolution of discs (starbursts induced by bar 
instabilities).  \cite{Bouwens1999} argue that all these
mechanisms could have been at work over the history of the Universe in forming 
bulges. \cite{Nakasato2003} have performed three-dimensional hydrodynamical 
N-body simulations of the formation of the Galactic bulge, and they find that 
two groups
of stars are present in the bulge, 60\% of the stars originate in
sub-galactic mergers, and 40\% come from the inner region of the disc. 
The two groups are chemically distinct. Their model is able to reproduce the
metallicity distribution and the velocity dispersion-metallicity relation.

The observational signatures of the bulge population are the following :
the bulge is certainly triaxial. The asymmetry is seen from COBE/DIRBE 
near-infrared data : the bulge is brighter at positive longitudes than at
negative ones. \cite{Dwek1995} and \cite{Freudenreich1998} have fitted 
parametric models to the dereddened DIRBE data, but excluding low latitude 
regions. \cite{Binney1997}, then \cite{Bissantz2002} have obtained non-parametric models from the same data.
The observed metallicity distribution of K giants is characterized by a wide
distribution, [Fe/H] ranging from -1.2 to +0.9 dex \citep{McWilliam1994}. \cite{Minniti1996} and \cite{Tiede1999} have shown a clear relation between the velocity dispersion and the 
metallicity but this result may be due to a mixture of populations (halo, 
bulge and maybe thick disc) in their fields at latitudes between -6 and -8.  
The bulge is rotationally supported, the projected mean rotation of the
bulge being of the order of $53.3 \pm 0.3$ km s$^{-1}$ kpc$^{-1}$ \citep{Tiede1999}.}

We here concentrate on the outer bulge, excluding 
the very central part of the Galaxy (inner bulge at longitude and 
latitude smaller than 1 deg) not yet well observationally constrained. 

As a first approximation, we
have started with the Dwek G2 density law \citep{Dwek1995} and have fitted
the parameters on near-infrared star counts from the DENIS survey. 
A detailed description of the process 
will be given elsewhere (Picaud et al., in preparation). Here we give the 
main results from this work.

The disc scale length is adjusted at the same time as the
bulge parameters. The data towards the inner Galaxy clearly show a coexistence
of both populations. However, since the Einasto formula for the disc accounts
for a possible hole in the middle of the disc, we take the opportunity to
constrain this parameter and check how the bulge and disc populations coexist
or are replaced by each other as a function of Galactic radius.

The density laws are given in table~3. The fitted parameters 
for the disc
are the scale length $h_{R_+}$ and the hole scale length $h_{R_-}$. For the bulge 
there are 8 fitted parameters (following Dwek notation):

\begin{itemize}
\item	orientation angles: $\alpha$ (angle between the bulge major axis
and the line perpendicular to the Sun - Galactic Center line ), 
$\beta$ (tilt angle between the bulge
plane and the Galactic plane) and $\gamma$ (roll angle around the bulge major
axis)
\item	scale-lengths $x_0$ on the major axis, $y_0$ and $z_0$ on minor axes
\item 	normalization (star density at the center) $N$ and cut-off
radius $R_c$.
\end{itemize}

In order to perform the Monte-Carlo simulations, the bulge luminosity 
function is taken from \cite{1997AJ....114.1531B} at solar 
metallicity and age of 10 Gyr. Alternatively Padova isochrones may be
used in the future. A Salpeter IMF has been assumed (see table~1). This
IMF slope seems to be well constrained for high mass and intermediate mass 
stars. This is however not the case at low masses, for which very deep
observations inside the bulge are needed but difficult to get because of
the crowding. From deep photometry obtained with the Hubble Space Telescope \cite{Holtzman1998} found a power-law mass function with a slope -2.2 for masses
higher than 0.7, though close to the Salpeter IMF. At lower masses they 
found indications of a break in the mass function,
as seen in other populations, occuring at about 0.5 to 0.7 \Msun. 
However, due to incompleteness and uncertain binary corrections, they do
not strongly constrain the value of the IMF slope at low masses.
If we would follow the general scheme of other populations, we would have 
taken either the
low mass slope of the disc (0.6) or that of the thick disc and stellar 
halo (-0.5). 
The constraints being still poor we have kept the one-slope Salpeter IMF
for the time being.

The DENIS survey \citep{Epchtein} has produced specific observations in the
bulge well suited for this analysis. We have used catalogues in J 
(1.25 $\mu m$) 
and $K_{s}$ (2.15 $\mu m$) bands from a set
of ``batches'', specific observations made in the directions of the bulge
(Simon et al, in preparation).
Contrary to the strips done in the
main survey, the batches are shorter in declination and easier to handle. 
About 100 windows of low extinction of 15'x15' between longitudes [-8$^\circ$, +12$^\circ$] 
and latitudes [-4$^\circ$; +4$^\circ$] were chosen using the
map of extinction from \cite{Schultheis}. The completeness limits were 
carefully established in each band. Star counts in K and J-K were simulated
and compared with the data
between magnitude  K 8.5 and 11.5 after extinction was fitted independently in
each window. 

The model goodness-of-fit is estimated through a maximum likelihood test.
In order to explore the 10 dimensional space of parameters, 
sets of parameters were chosen by Monte Carlo selections, first from a
uniform distribution, then
from a Gaussian distribution around the maxima of likelihood.
Fifty independent fits were made to test the robustness of the
results.

The best fit parameters (mean of the 50 trials and dispersions about the mean) 
are given in table~5.
\begin{table*}
{\centering 
\caption{Bulge fitted parameters. Values from  \protect\cite{Dwek1995} are mean values
obtained from the 2.2 $\mu m$ DIRBE data for their G2 density law. $\alpha$, $\beta$ and $\gamma$ are orientation angles following the \cite{Dwek1995} notation. $x_0$,$y_0$,  $z_0$ are
the scale lengths in the three axes. $R_c$ is the cutoff radius and $N$ the
central number density. The values of $\beta$ and Rc are not adjusted in the 
case of the  \protect\cite{Dwek1995} fitting process.}
\begin{tabular}{llllll}
\hline
 Parameter & Dwek et al. value & Error &Our fitting value & Error & Unit\\
$\alpha$ &68.5 &4.8 &78.9$^\circ$ & 0.7 &$^\circ$\\
$\beta$ &0 & - &3.5$^\circ$ & 0.2 & $^\circ$\\
$\gamma$ &90.7 &1.5 &91.3$^\circ$ & 1.1 & $^\circ$\\
$x_0$ &2.03 &0.23 &1.59 & 0.12 &kpc\\
$y_0$ &0.470 &0.010 &0.424 & 0.015 &kpc\\
$z_0$ &0.580 &0.010 &0.424 & 0.015 &kpc\\
$R_c$ &2.4 &- &2.54 & 0.13 &kpc\\
$N$ & - & - & 13.70  &0.7 & star.pc$^{-3}$\\
\hline
\end{tabular}\par}
\end{table*}

	{ Thus, we obtain a very oblate outer bulge (sometimes called a bar 
in the literature), almost in the
Galactic Plane with an angle from the Sun - Galactic Center line of about
10 degrees, well in agreement with the result from \cite{lopez2000} who
found an angle of 12 degrees from independent star count data. 
{ The axis ratios are in good agreement with the \cite{Dwek1995} 
determination but they found a slightly higher angle, although 
less constrained, 
between the major axis and the line of sight to the Galactic center (20$^{\circ}\pm 10^{\circ}$) compared to ours 
(11.1$^{\circ}\pm0.7^\circ$). The good accuracy we achieved comes from
the use of star counts, more
sensitive to lower luminosity stars covering a larger depth inside the bulge
and with a higher density,
rather than integrated luminosities, dominated by the brightest stars, which
may be biased by the deprojection.

The overall mass in this outer bulge is 
$2.03 \pm 0.26~\times~10^{10}~\Msun$. It includes all populations, even
white dwarfs which contribute to about 26\%. It does not include the inner
bulge or the central black hole. This value is slightly higher than the
one determined by \cite{Dwek1995} from COBE data $1.3~\times~10^{10}$\Msun, but
the latter is photometrically determined and does not account for white
dwarfs. 

The overall fit of the central region leads to include a noticeable hole
in the thin disc, which is very common in barred galaxies. The resulting disc 
scale length is $h_{R_+} = 2.53\pm 0.11$ kpc and the hole scale length
$h_{R_-} = 1.32 \pm 0.14$ kpc.}}

The bar is considered as a separate population from the bulge, extending
further away (up to l=27$^\circ$ as seen by \cite{lopez2000}). It is being
studied and is discussed elsewhere \citep{Picaud2003}.

\subsection{White dwarfs}

{ White dwarfs (WD) are the latest stage of stellar evolution. Their density is a function of the past star formation history and the initial
mass function. Their luminosity function depends on the assumed cooling tracks
and physical properties. In the modeling we assume all white dwarfs to be DA
and use evolutionary tracks and atmosphere models from
\cite{Bergeron1995}, complemented by 
\cite{Chabrier1999} for the very cool end, applicable to the halo. 
The former have been extensively
compared with observations and have been shown to be reliable. The latter 
models still have
to be checked, as very few ancient white dwarfs are known yet. However,
they include the best physical constraints and are supposed to be good starting points for simulating ancient white dwarfs.}

\subsubsection{Thin disc}
{ The luminosity function (LF) of disc white dwarfs is known from systematic
searches in the solar neighbourhood. A good preliminary LF was obtained by \cite{Liebert1988}}. More recent determinations 
give larger local densities due to new detections of cool white dwarfs: 
\cite{2002ApJ...571..512H} found $5.5 \pm 0.8 \times 10^{-3}$ star~pc$^{-3}$, 
while \cite{1998ApJ...497..294L} and \cite{Ruiz2001} found
a lower limit of $5.6 \times 10^{-3}$ star~pc$^{-3}$, and \cite{Knox1999} 
$4.16 \times 10^{-3}$ star~pc$^{-3}$. It should be noted that these values
depend a lot on the number of faintest white dwarfs detected at M$_V>16$,
which are only 3, and the way the luminosity function is binned. 
\cite{Wood1998} estimated that the accuracy of the
white dwarf density is about 50\% on a sample of 50 objects (the number
of white dwarfs known in the local sphere of 13 pc is 53).
 
{ To avoid Poisson noise from observational data, we have
adopted a luminosity function based on the \cite{Wood1992} model for a disc
of 10 Gyr, with a constant star formation rate, coherent with our evolutionary model}. This is well in agreement
with the bright DA luminosity function from \cite{Liebert1988} and is slightly
higher in the coolest part of the observed white dwarf luminosity function, 
allowing for incompleteness in the local sample.
The total local density is then  6.6 $\times 10^{-3}$ star~pc$^{-3}$ and 
the mass density 3.96 $\times 10^{-3}$ $\Msun$ pc$^{-3}$.

\subsubsection{Thick disc}

The thick disc white dwarf population has been 
computed by Garcia-Berro (private communication and \cite{Garcia-Berro1999})
assuming a Salpeter IMF. Very few thick disc white dwarfs have yet been
identified in the field. Hence, we have tentatively 
normalized the theoretical LF
 by fitting model predictions on the \cite{Oppenheimer2001} 
sample of high velocity white dwarfs. \cite{Reyle2001b} have shown that
most of the Oppenheimer WD belong to the thick disc populations, as
suggested by their kinematical properties. It gives 
a density of white dwarfs relative to 
the red dwarfs of 20\% that is 5$\times 10^{-4}$ stars pc$^{-3}$. 

\subsubsection{Halo white dwarfs}

Due to the microlensing results towards the Magellanic Clouds, the hypothesis
that a significant number of ancient white dwarfs are present in the halo is
seriously considered. The local contribution of this faint population in the
solar neighbourhood is detectable and several studies have been conducted in
order to find them. Thus, this hypothetical population has been tentatively
included in the model.

The current estimation of machos density 
from microlensing events towards the 
Magellanic Clouds \citep{Lasserre2000,Alcock2000} has an upper limit of
about 20\% of dark matter halo made of machos. \cite{Freese1997} have shown
the improbability of the machos to be red or brown dwarfs, leaving the
possibility for ancient white dwarfs to be the lenses. These results
have stimulated the search for the local counterpart of the halo
baryonic dark matter in the form of white dwarfs. 
However, the results give a quite smaller local density :
\cite{Goldman2002} have shown that the local density of halo white 
dwarfs cannot
exceed 5\% of the dark halo and \cite{Oppenheimer2001} give a lower limit 
of 3\%,
although we do not think, from their kinematics, that all Oppenheimer 
objects are 
really halo white dwarfs \citep{Reyle2001b}. \cite{Mendez2002}
has recently found a population of old white dwarfs in a proper motion survey
near the globular cluster NGC6397. Among his 7 peculiar white dwarfs he claims
to have a halo white dwarf and 6 thick disc white dwarfs. If he is right, then
the number of thick disc white dwarfs is huge compared with the number 
generally assumed
and they have (for 4 among 6) a surprising rotational velocity higher than
the LSR, while one expects a significantly smaller rotational velocity for
the thick disc population. The identification of these stars as belonging to
the thick disc is then problematic. One should consider rather the 
possibility of them being disc white dwarfs which have undergone a kick off
in a mass-transfer binary, as proposed by \cite{Bergeron2003}. 
Considering the Mendez's halo white dwarf 
candidate, the halo white dwarf mass density inferred by this object
would be about half the density of the dark halo 
(4.25 $\pm$ 2.8 \Msun pc$^{-3}$) if this star is a halo object. Its proper
motion is still compatible with thick disc kinematics.

Another recent search for high proper motion objects in the 
VIRMOS survey \citep{Lefevre1998} on 0.94~square degree
adding to a 0.16~square degree in the SA57 field obtained also at the CFHT
produces no detection of high proper motion white
dwarf in selection limits, where a 100\% white dwarf halo aged 14 Gyr
would place between 12.6 and 20.4 detectable
objects (depending on the adopted halo IMF). This sets at the 95\%
probability level the upper limit of the halo fraction in the form of
halo white dwarfs between 24\% and 15\% (Cr\'ez\'e et al, in preparation).
The thick disc white dwarf population seen by Mendez has not been seen
in this Virmos field.

{ \cite{Chabrier1996} have reconsidered the hydrogen white dwarf cooling 
sequence and recomputed their model atmosphere and color-magnitude diagrams. 
In order to build their luminosity function and detection probabilities, they 
considered two different IMFs \citep{Chabrier1999}. They are mainly 
constrained by the nucleosynthesis
which places limits on the amount of high mass elements rejected in the 
interstellar medium at early ages and by the fact that none were yet observed
in white dwarf systematic searches. Of course there are not yet any direct
observational constraints on the true IMF of these first stars. Hence we have
alternatively used both IMF to compute dedicated simulations. The local 
normalization is assumed to be 2\% as a low conservative value compatible with
ancient white dwarf searches. The mean white dwarf mass is about 
0.8 \Msun for Chabrier's IMF1, and 0.7 \Msun for IMF2. The local number
density is deduced from these numbers and the local density of the dark matter
halo as determined by the dynamical constraints (sect. 2.1.1). We end up
with a local density of $2.5~\times~10^{-4}$~star~pc$^{-3}$ for IMF1 and 
$2.8~\times~10^{-4}$~star~pc$^{-3}$ for IMF2.}

\subsection{Interstellar extinction}

{ The interstellar matter is mostly concentrated in the Galactic plane. Hence
the intervening material can be modeled, at medium and high latitudes, 
as a double exponential or
an Einasto law, similar to the thin disc of youngest stars. However, 
when comparing model simulations with
data in the Galactic plane, this smooth model is representative of the
mean extinction only, and does not account for small scale variations of 
the extinction. 
At very low latitudes, it is better to apply, on a case by case basis, 
alternative distributions
determined from observed colour-magnitude diagrams or from independent
studies. \cite{Drimmel2001} have built a 3D extinction model, in which the
distribution of extinction clouds in a thin disc and spiral arms is forced
with the far-infrared emission due to the dust, as observed by COBE, or as 
determined by \cite{Schlegel}. An overall analysis of the near infrared 
star counts from recent all-sky surveys should provide in the near future 
a more accurate extinction distribution due to dust in the Galactic disc.}

The reddening in each photometric band is obtained from the extinction law 
of \cite{Mathis1990}. 
{ \cite{Cardelli} have shown that their extinction law, very closed to
the latter, is well reliable over
the wavelength range 0.125 $\mu m < \lambda < 3.5\mu m$ and is applicable to both
diffuse and dense regions of the interstellar medium, except for very 
dense clouds. We adopt a value of
the commonly used parameter R$_V$ = A(V)/E(B-V) = 3.1. This value may be 
inappropriate in some particular lines of sight in highly extincted clouds,
which we do not intend to model in detail.
Moreover the effect of the choice of R$_V$ is negligible at $\lambda~>~0.7 \mu m$.}

\section{Model simulations and applications}

\subsection{Simulations}

Using the inputs described above, we compute the number of stars of a 
given age, type, 
effective temperature, and absolute magnitude, at any place in the Galaxy. 
The assumed sun position
is at a galactocentric radius of 8.5 kpc (the IAU value) and at 15 pc above the
plane \citep{Freudenreich1998,Hammersley95,Drimmel2001}.
From the expected density of
a given type of stars, a series of Monte-Carlo drawings are performed to make
a sample of stars, the number of which is compatible with the expectation and
includes the Poisson noise. Thus, when producing a series of catalogues
in the same theoretical conditions one obtains variations in the number of
stars compatible with the Poisson statistics.

Apparent magnitudes and colours are then computed star by star
for a given set of filters with
the above mentioned model atmospheres,
reddened as a function of intervening extinction on the line of sight.
The photometric bands currently available are the Johnson-Cousins system
(UBVRIJHKL) and the GAIA G magnitude. Predictions at 7 and 15 microns 
for ISO bands, 12 and 25 microns IRAS bands are being tested, as well as
 the Sloan photometric system.
Photometric errors are included on each photometric band
independently, the 
error being a function of the magnitude (as usually observed). This function
can be a third degree polynomial, or an exponential as in equation~2. 

\begin{equation}
\sigma = A + exp(C\times m - B) 
\end{equation}

where m is the magnitude in the band considered, and A,B,C are fitted to the 
set of data with which the
model predictions are going to be compared.
The polynomial function generally gives a realistic fit to photographic 
photometry while the exponential reproduces very well typical CCD photometric 
errors.

The astrometry is computed using the kinematical parameters summarized in 
table~4.
Astrometric errors are assumed to follow a function of the magnitude, as the
photometry, either polynomial or exponential.
Radial velocities are computed in the same way. If needed, parallaxes can 
also be estimated, assuming that the proper error function is known.

Finally, a catalogue (or a statistics if needed) of the expected
star distribution in any given direction of observation is created. 
The output parameters
for each simulated star include the intrinsic parameters (absolute magnitude,
effective temperature, gravity, age, metallicity [Fe/H], U,V,W velocities computed without errors) as well as observational characteristics (apparent 
magnitude, colours, proper motion, radial velocity, parallax), distance 
to the sun and interstellar extinction.

\subsection{Basic applications}

{ The model is able to produce star counts in any given direction but the 
Galactic center (the inner bulge being not yet included). It produces
alternatively a catalogue of simulated stars or statistics on any given
observable or intrinsic property.

The model has already been widely used (either in its previous version or
in the present one) mainly for the following purposes :

\begin{itemize}
\item Predictions of star counts in different photometric bands 
from U to K in different
directions to be compared with observational data: examples of recent 
comparisons
between model and observed star counts can be found in \cite{Goldman2002},\cite{Ojha2001},\cite{Robin2000},\cite{Reyle2001a} and \cite{Castellani2001} in visible bands; 
\cite{Krause2003} in the I band; 
\cite{Ruphy1996},\cite{Persi1999}, \cite{Persi2001} and \cite{Reyle2001a} in the near infrared. 

\item Predictions in the X domain:
\cite{Guillout1996} have extended the predictions 
of the previous model version to ROSAT bands. It has been used to compute the 
contributions of Galactic stars to the diffuse X-ray background
calculated for ROSAT PSPC energy bands \citep{Kuntz2001}.

\item Predictions of kinematics and comparison with observational data \citep{Bienayme1992,Chareton1993,Ojha1999,Rapaport2001,Soubiran2003}.

\item Predictions for use in preparing observations. The model allows
to make simulations in different conditions, to test the limiting
magnitude to be reached for answering a given question and to choose the
most efficient filters (see also the work of
\cite{Lelouarn2000} for studying adaptative optics possibilities with 
guide stars).

\item Determinations of extinction in the Galactic plane \citep{Zagury1999}.
We have also shown that the model can be used to map the extinction, just 
by playing with the extinction distribution along a line of sight 
introduced in a simulation. This will be used in the future to build a
tri-dimensional model of the interstellar extinction.

\item Estimations of expected contamination of Galactic star in front of
clusters, molecular clouds, or nearby resolved galaxies (see for
example \cite{Pierre1987,Cananzi1992,Thoraval1996,Deeg1998,Cambresy1998,Vuong2001,Rejkuba2001,Rejkuba2002,Moraux2001}.

\item Simulations of proper motion selected samples to correct 
kinematically biased selections \citep{Reyle2001b}.

\end{itemize}

The wide variety of simulations which can be done allow to obtain a wide set
of different type of constraints for the model. Hence it furnishes a wide 
variety of ways to constrain the scenario of Galactic formation.

In order to test model hypothesis, predictions have been abundantly 
compared with star counts in many directions,
photometric bands from B to K, in magnitudes, colours and proper motions.

We give here just a sample of model predictions compared with observational 
data in magnitudes V and K and a colour/proper motion distribution.
Figures~4 to 7 show star count predictions in the V band as a function
of longitude and latitude. We have added a few observational measurements
from the literature. The agreement between
model and data is satisfying and gives confidence for using this model for
interpolations.  It should be noted
here that the extinction has not been adjusted in the Galactic plane, hence
predictions given at b=0$^\circ$ may be inadequate.
Figures~8 to 10 show star count predictions at 2.2 $\mu$m
and are compared with a few samples from the DENIS survey. In figures~5 
and 8 we have
not given the expected star counts at the Galactic center, as the inner
bulge is not modeled.}

In figure~6 and 9, the numbers are given for longitudes 90 and 270,
for both quadrants. The counts vary slightly between these two directions
at V$>$12 and $|b|<45$ because of the warp.
In the anticenter direction, as seen in figure~7, the effect of the
flare is that the counts at latitude 10$^\circ$ are at the 
same level as at latitude~0.

Figure~11 gives the contributions of various
spectral types as a function of V magnitude in the direction l=90$^\circ$
and b=45$^\circ$, representative of what happens at medium latitudes. Giants
dominate the counts up to magnitude 11 in V, where main sequence stars 
start
to be more numerous. The limiting magnitude between these two regimes moves
to fainter magnitudes when the latitude goes down.

Figure~12 shows a comparison between observed and modeled colour-proper
motion diagram towards l=$3^\circ$ and b=$47^\circ$ in the magnitude
range $12<V<18$. Data are from 
\cite{Ojha1996}. The model reproduces well the overall distribution, 
the mean and dispersion in proper motion, and the colour distribution: the 
red part
is dominated by the disc population, the blue one by thick disc and 
spheroid stars. 

{
More comparisons between model predictions and observational 
data can be 
found in previous published papers. Colour histograms simulated and compared
with observations can be found in \cite{Robin2000} in 15 directions.
In \cite{Reyle2001b} we show a velocity diagram made from a simulation of 
a proper motion selected sample,
to be compared with the \cite{Oppenheimer2001} sample of probable white dwarfs.}

\subsection{Bayesian classification}
Apart from common predictions, star counts and catalogues of simulated stars, 
the model predictions may also be useful for estimating Bayesian 
probabilities: given a set of 
observables and observational errors, by suitable simulations of the direction
of interest, the model can establish a probability classification. This 
classification
is independent from any spectral classification: assuming that the 
model reproduces correctly the stellar content of the Galaxy, it allows
to perform an estimate of the probable class, metallicity or population 
of an observed star, using different types of observables such as 
magnitudes, colours, proper motions, radial velocities, etc.

A previous use of this Bayesian classification can be found in 
\cite{Reyle2002}. In a proper motion selection sample of nearby dwarfs,
the combined use of magnitude, colours and proper motions allowed to determine
the probable membership of observed stars, to deduce a metallicity and a
distance (as the distance estimator is metallicity dependent). 
We think that this approach can be used on many occasions as a complementary
classification to usual photometric or spectral classification{, in particular
in the framework of Virtual Observatories}. The limitation
of this classification is the adequacy of the model to the data. One has
to verify in each case that the model reproduces well the distribution of
available observables { before relying on such classification.}

\section{Conclusion and future plans for model developments}

The model may suffer from misfits in fields close
to the Galactic plane due to the crude model of extinction, which does 
not account for high or irregular absorbing clouds and the absence
of spiral arm modelisation. { While we do not intend to model any stellar 
and interstellar fluctuations in the Galactic plane, at least an estimate
of the amount of extinction, and its variation along the line of sight, is 
needed to interpret observations in the disc, even in the infrared. We intend
to participate to the challenge of getting a reliable 3D model of extinction,
such as that the one of \cite{Drimmel2001}, but constrained by observed
star counts, which, we think, are the more reliable way of determining the
distance of the absorption features.}

{At the moment, the model is well constrained by star counts at magnitudes
between 12 and 24. At both ends, deviations are found and should be
studied in more detail.}
Extrapolations at very faint magnitudes are still risky since
the predictions strongly depend on the assumed IMF at low masses which 
are still very uncertain. 
Figure~4 shows model predictions and observed star counts towards
the Galactic pole in the V band. The overall agreement is very good up to
V=22. At deeper magnitudes the agreement would be
better by adopting
a higher IMF slope for the spheroid ($\alpha=2$, dotted dashed line), 
compared with the standard value of 0.5 (solid line).
But we have only one data set to rely on at these magnitudes and we 
suspect the observations  
to be contaminated by galaxies at V$>22$.
The determination of the spheroid IMF at low masses relies
on a few very deep star counts: space based counts are scarce and have 
poor statistics, while ground based ones suffer from contamination by 
galaxies. This point will certainly be addressed by ongoing wide field
deep surveys, such as the coming CFHTLS, VST or VISTA surveys, and from
the ACS camera of the HST.

Nevertheless the present version of the Besan\c{c}on model of the Galaxy
(available at http://www.obs-besancon.fr/modele/model2003.html)
is well suited for testing Galactic structure parameters and evolution 
scenarios,
and is useful for accurate photometric and astrometric 
predictions in different filters from
the visible to the near-infrared. We plan to provide calibrations for 
the SDSS system, and for the future GAIA photometric system, when chosen.
The source counts in ROSAT bands from \cite{Guillout1996} will be adapted
for XMM satellite bandpasses. We have also computed
magnitudes in some ISOCAM filters which will be used in the near
future for interpreting the ISOGAL survey \citep{Omont2000}.

The mass model can be used to compute optical depths towards the
inner Galaxy and the Magellanic Clouds, to be compared with constraints from
microlensing experiments. This is also one of our aims to use these results 
to analyse self-consistently the measured optical depths fully 
taking into account
the current knowledge on Galactic structure and populations.

\begin{figure*}
\label{pole}
\begin{center}
\begingroup%
  \makeatletter%
  \newcommand{\GNUPLOTspecial}{%
    \@sanitize\catcode`\%=14\relax\special}%
  \setlength{\unitlength}{0.1bp}%
\begin{picture}(4320,2592)(0,0)%
\special{psfile=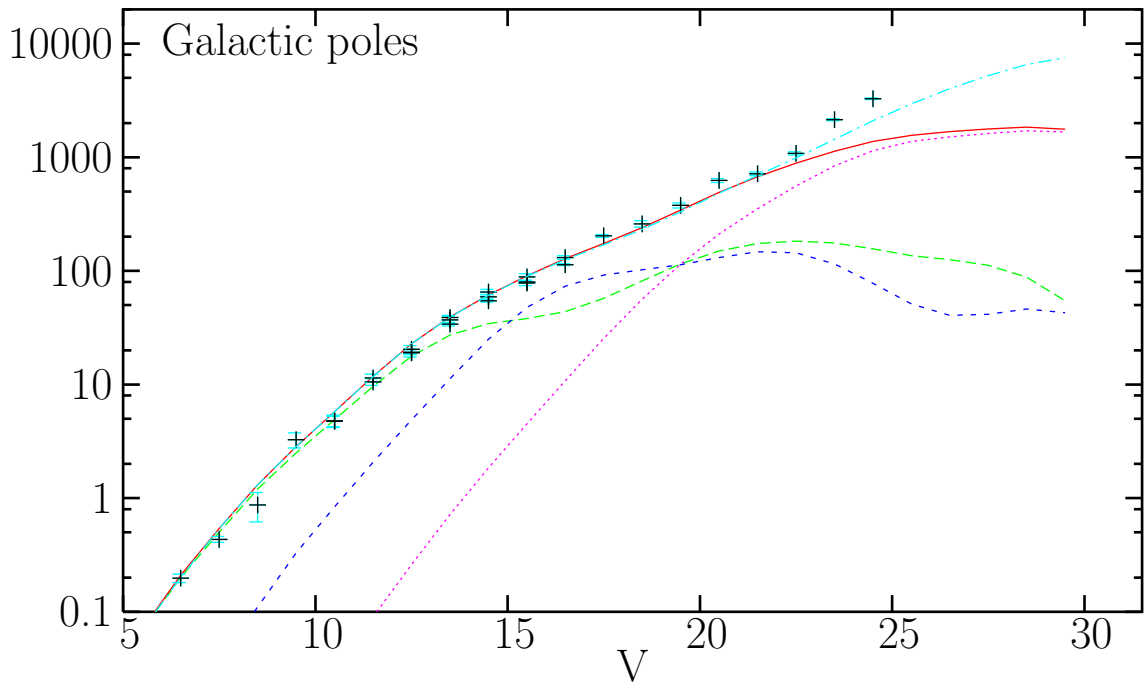 llx=0 lly=0 urx=432 ury=259 rwi=4320}
\fontsize{16}{\baselineskip}\selectfont
\put(505,2383){\makebox(0,0)[l]{Galactic poles}}%
\put(2280,40){\makebox(0,0){V}}%
\put(3983,160){\makebox(0,0){ 30}}%
\put(3258,160){\makebox(0,0){ 25}}%
\put(2534,160){\makebox(0,0){ 20}}%
\put(1809,160){\makebox(0,0){ 15}}%
\put(1085,160){\makebox(0,0){ 10}}%
\put(360,160){\makebox(0,0){ 5}}%
\put(320,2383){\makebox(0,0)[r]{ 10000}}%
\put(320,1954){\makebox(0,0)[r]{ 1000}}%
\put(320,1526){\makebox(0,0)[r]{ 100}}%
\put(320,1097){\makebox(0,0)[r]{ 10}}%
\put(320,669){\makebox(0,0)[r]{ 1}}%
\put(320,240){\makebox(0,0)[r]{ 0.1}}%
\end{picture}%
\endgroup
 
\caption{Star count predictions in the V band at the Galactic pole. 
Solid line : total with
a spheroid IMF slope of $\alpha=0.5$, long dashed: disc, short dashed:
thick disc, dotted: spheroid, dotted dashed: total with
a spheroid IMF slope of $\alpha=2$. Various observed counts are 
indicated as crosses (error bars are the Poisson noise) from Simbad
data base, 
\protect\cite{Gilmore1985},  \protect\cite{Bok1964},
\protect\cite{Chiu1980}, 
 \protect\cite{Yoshii1987} and Cr\'ez\'e et al. in preparation.}
\end{center}
\end{figure*}

\begin{figure*}
\begin{center}
\begingroup%
  \makeatletter%
  \newcommand{\GNUPLOTspecial}{%
    \@sanitize\catcode`\%=14\relax\special}%
  \setlength{\unitlength}{0.1bp}%
\begin{picture}(4320,2592)(0,0)%
\special{psfile=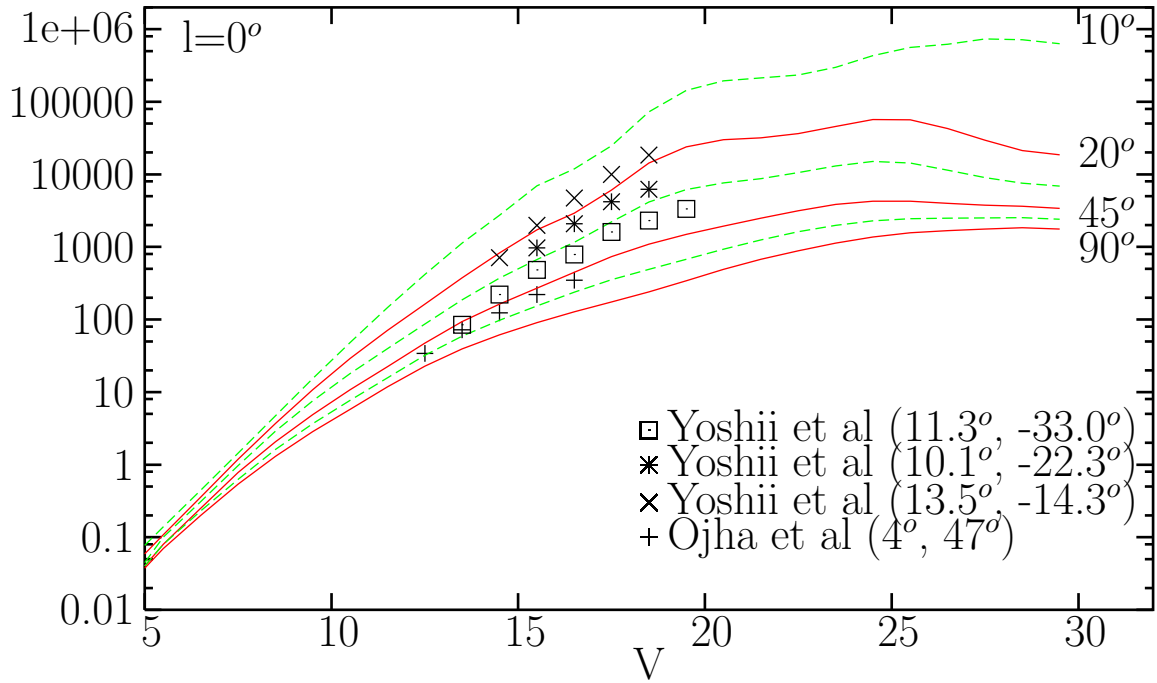 llx=0 lly=0 urx=432 ury=259 rwi=4320}
\fontsize{16}{\baselineskip}\selectfont
\put(2370,918){\makebox(0,0)[l]{Yoshii et al (11.3$^o$, -33.0$^o$)}}%
\put(2370,787){\makebox(0,0)[l]{Yoshii et al (10.1$^o$, -22.3$^o$)}}%
\put(2370,644){\makebox(0,0)[l]{Yoshii et al (13.5$^o$, -14.3$^o$)}}%
\put(2370,514){\makebox(0,0)[l]{Ojha et al (4$^o$, 47$^o$)}}%
\put(3919,1609){\makebox(0,0)[l]{90$^o$}}%
\put(3919,1739){\makebox(0,0)[l]{45$^o$}}%
\put(3919,1965){\makebox(0,0)[l]{20$^o$}}%
\put(3919,2430){\makebox(0,0)[l]{10$^o$}}%
\put(541,2404){\makebox(0,0)[l]{l=0$^o$}}%
\put(2300,40){\makebox(0,0){V}}%
\put(3919,160){\makebox(0,0){ 30}}%
\put(3215,160){\makebox(0,0){ 25}}%
\put(2511,160){\makebox(0,0){ 20}}%
\put(1807,160){\makebox(0,0){ 15}}%
\put(1104,160){\makebox(0,0){ 10}}%
\put(400,160){\makebox(0,0){ 5}}%
\put(360,2430){\makebox(0,0)[r]{ 1e+06}}%
\put(360,2156){\makebox(0,0)[r]{ 100000}}%
\put(360,1882){\makebox(0,0)[r]{ 10000}}%
\put(360,1609){\makebox(0,0)[r]{ 1000}}%
\put(360,1335){\makebox(0,0)[r]{ 100}}%
\put(360,1061){\makebox(0,0)[r]{ 10}}%
\put(360,787){\makebox(0,0)[r]{ 1}}%
\put(360,514){\makebox(0,0)[r]{ 0.1}}%
\put(360,240){\makebox(0,0)[r]{ 0.01}}%
\end{picture}%
\endgroup
 
\small
\caption{Star count predictions (stars per magnitude and per 
square degree) in the V band at l=0$^\circ$, for latitudes 10$^\circ$ to 90$^\circ$
from top to bottom (20$^\circ$,45$^\circ$ and 90$^\circ$ with solid lines, 10$^\circ$,30$^\circ$,60$^\circ$ with dashed 
lines). Data are from \protect\cite{Ojha1994a} and  \protect\cite{Yoshii1989}.}
\end{center}
\end{figure*}
\begin{figure*}
\begin{center}
\begingroup%
  \makeatletter%
  \newcommand{\GNUPLOTspecial}{%
    \@sanitize\catcode`\%=14\relax\special}%
  \setlength{\unitlength}{0.1bp}%
\begin{picture}(4320,2592)(0,0)%
\special{psfile=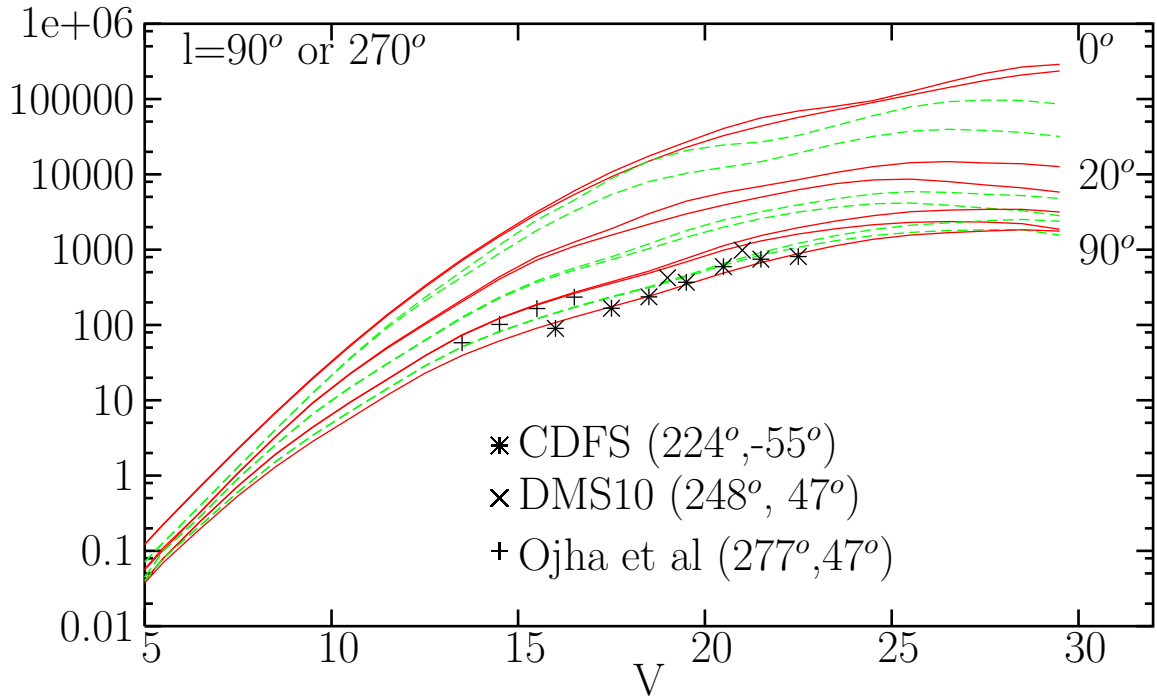 llx=0 lly=0 urx=432 ury=259 rwi=4320}
\fontsize{16}{\baselineskip}\selectfont
\put(1807,921){\makebox(0,0)[l]{CDFS (224$^o$,-55$^o$)}}%
\put(1807,723){\makebox(0,0)[l]{DMS10 (248$^o$, 47$^o$)}}%
\put(1807,496){\makebox(0,0)[l]{Ojha et al (277$^o$,47$^o$)}}%
\put(3919,1660){\makebox(0,0)[l]{90$^o$}}%
\put(3919,1944){\makebox(0,0)[l]{20$^o$}}%
\put(3919,2420){\makebox(0,0)[l]{0$^o$}}%
\put(541,2420){\makebox(0,0)[l]{l=90$^o$ or 270$^o$}}%
\put(2300,40){\makebox(0,0){V}}%
\put(3919,160){\makebox(0,0){ 30}}%
\put(3215,160){\makebox(0,0){ 25}}%
\put(2511,160){\makebox(0,0){ 20}}%
\put(1807,160){\makebox(0,0){ 15}}%
\put(1104,160){\makebox(0,0){ 10}}%
\put(400,160){\makebox(0,0){ 5}}%
\put(360,2512){\makebox(0,0)[r]{ 1e+06}}%
\put(360,2228){\makebox(0,0)[r]{ 100000}}%
\put(360,1944){\makebox(0,0)[r]{ 10000}}%
\put(360,1660){\makebox(0,0)[r]{ 1000}}%
\put(360,1376){\makebox(0,0)[r]{ 100}}%
\put(360,1092){\makebox(0,0)[r]{ 10}}%
\put(360,808){\makebox(0,0)[r]{ 1}}%
\put(360,524){\makebox(0,0)[r]{ 0.1}}%
\put(360,240){\makebox(0,0)[r]{ 0.01}}%
\end{picture}%
\endgroup
 
\caption{Star count predictions (stars per magnitude and per 
square degree) in the V band at l=90$^\circ$ or 270$^\circ$, for
latitudes 10$^\circ$ to 90$^\circ$
from top to bottom (0$^\circ$,20$^\circ$,45$^\circ$ and 90$^\circ$
with 
solid lines, 10$^\circ$,30$^\circ$,60$^\circ$ with dashed 
lines). For each latitude the bifurcation
is due to the warp, the highest part being for l=90, b$>$0 or l=270,
b$<$0. 
Data are from  \protect\cite{Ojha1996},
the DMS survey  \protect\cite{Hall1996ApJS} and in the Chandra Deep 
Field South from  \protect\cite{Groenewegen2002} 
(longitudes and 
latitudes of the data are in parentheses).}
\end{center}
\end{figure*}
\begin{figure*}
\begin{center}
\begingroup%
  \makeatletter%
  \newcommand{\GNUPLOTspecial}{%
    \@sanitize\catcode`\%=14\relax\special}%
  \setlength{\unitlength}{0.1bp}%
\begin{picture}(4320,2592)(0,0)%
\special{psfile=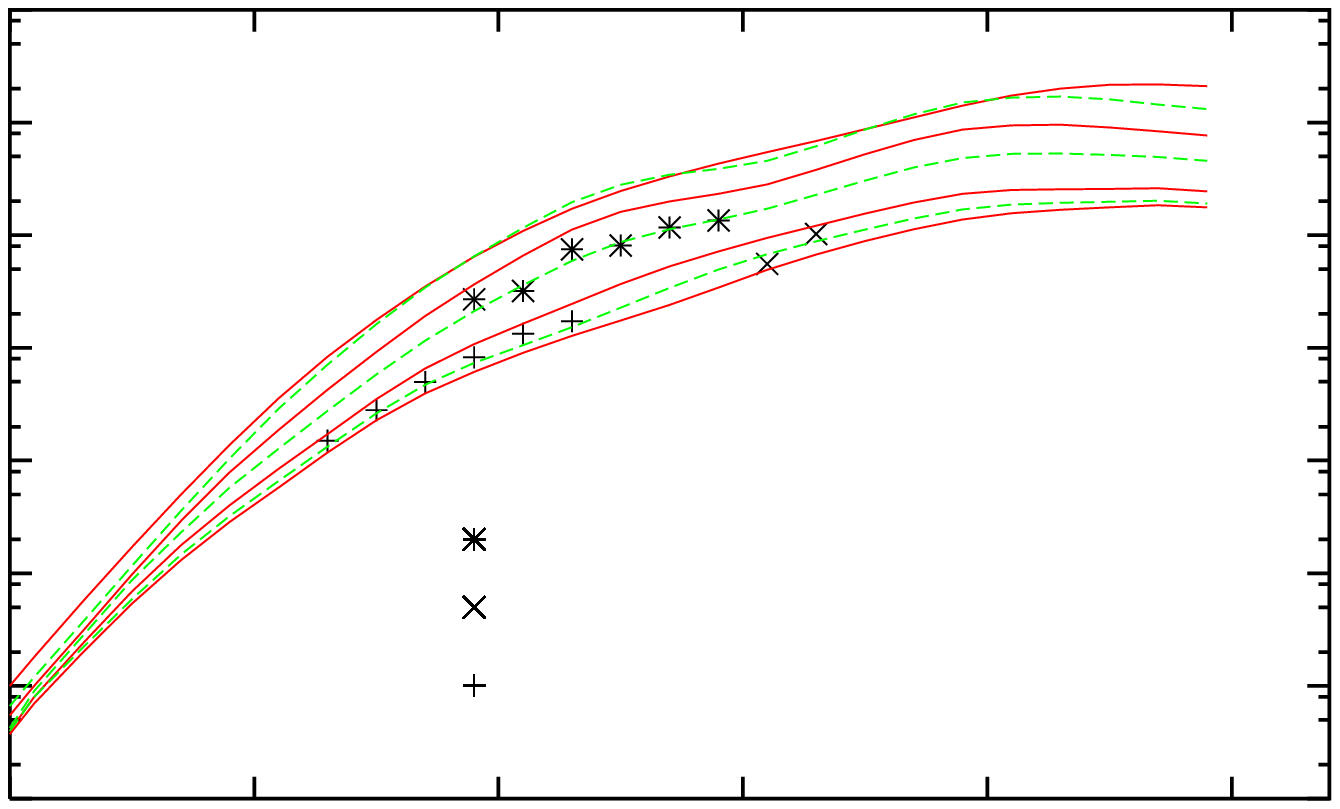 llx=0 lly=0 urx=432 ury=259 rwi=4320}
\fontsize{16}{\baselineskip}\selectfont
\put(1807,987){\makebox(0,0)[l]{Chiu (189$^o$,21$^o$)}}%
\put(1807,791){\makebox(0,0)[l]{Borra et al (197$^o$,38$^o$)}}%
\put(1807,565){\makebox(0,0)[l]{Ojha et al (168$^o$,48$^o$)}}%
\put(3919,1863){\makebox(0,0)[l]{90$^o$}}%
\put(3919,1992){\makebox(0,0)[l]{45$^o$}}%
\put(3919,2137){\makebox(0,0)[l]{20$^o$}}%
\put(3919,2285){\makebox(0,0)[l]{0$^o$}}%
\put(541,2409){\makebox(0,0)[l]{l=180$^o$}}%
\put(2300,40){\makebox(0,0){V}}%
\put(3919,160){\makebox(0,0){ 30}}%
\put(3215,160){\makebox(0,0){ 25}}%
\put(2511,160){\makebox(0,0){ 20}}%
\put(1807,160){\makebox(0,0){ 15}}%
\put(1104,160){\makebox(0,0){ 10}}%
\put(400,160){\makebox(0,0){ 5}}%
\put(360,2512){\makebox(0,0)[r]{ 100000}}%
\put(360,2187){\makebox(0,0)[r]{ 10000}}%
\put(360,1863){\makebox(0,0)[r]{ 1000}}%
\put(360,1538){\makebox(0,0)[r]{ 100}}%
\put(360,1214){\makebox(0,0)[r]{ 10}}%
\put(360,889){\makebox(0,0)[r]{ 1}}%
\put(360,565){\makebox(0,0)[r]{ 0.1}}%
\put(360,240){\makebox(0,0)[r]{ 0.01}}%
\end{picture}%
\endgroup
 
\caption{Same as figure~5 but for l=180. Data are from  \protect\cite{Ojha1994b},
 \protect\cite{Borra86} and  \protect\cite{Chiu1980} in indicated directions.}
\end{center}
\end{figure*}

\begin{figure*}
\begin{center}
\begingroup%
  \makeatletter%
  \newcommand{\GNUPLOTspecial}{%
    \@sanitize\catcode`\%=14\relax\special}%
  \setlength{\unitlength}{0.1bp}%
\begin{picture}(4320,2592)(0,0)%
\special{psfile=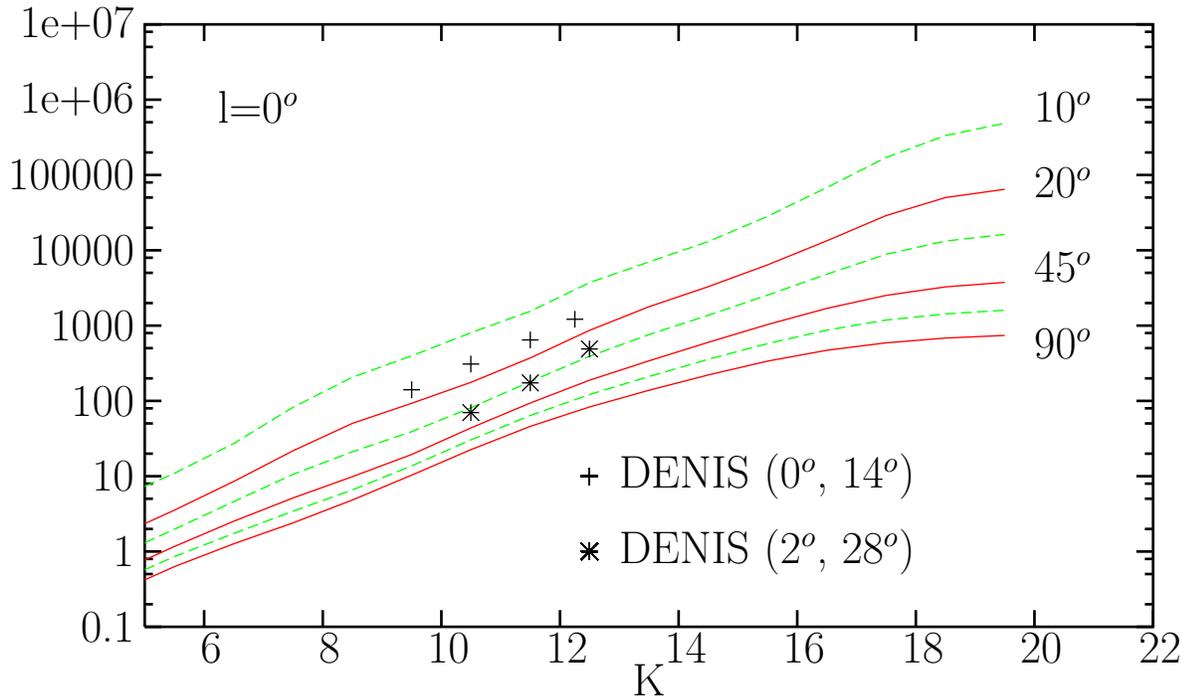 llx=0 lly=0 urx=432 ury=259 rwi=4320}
\fontsize{16}{\baselineskip}\selectfont
\put(2188,524){\makebox(0,0)[l]{DENIS (2$^o$, 28$^o$)}}%
\put(2188,808){\makebox(0,0)[l]{DENIS (0$^o$, 14$^o$)}}%
\put(3753,1309){\makebox(0,0)[l]{90$^o$}}%
\put(3753,1598){\makebox(0,0)[l]{45$^o$}}%
\put(3753,1916){\makebox(0,0)[l]{20$^o$}}%
\put(3753,2202){\makebox(0,0)[l]{10$^o$}}%
\put(624,2202){\makebox(0,0)[l]{ l=0$^o$}}%
\put(2300,40){\makebox(0,0){K}}%
\put(4200,160){\makebox(0,0){ 22}}%
\put(3753,160){\makebox(0,0){ 20}}%
\put(3306,160){\makebox(0,0){ 18}}%
\put(2859,160){\makebox(0,0){ 16}}%
\put(2412,160){\makebox(0,0){ 14}}%
\put(1965,160){\makebox(0,0){ 12}}%
\put(1518,160){\makebox(0,0){ 10}}%
\put(1071,160){\makebox(0,0){ 8}}%
\put(624,160){\makebox(0,0){ 6}}%
\put(360,2512){\makebox(0,0)[r]{ 1e+07}}%
\put(360,2228){\makebox(0,0)[r]{ 1e+06}}%
\put(360,1944){\makebox(0,0)[r]{ 100000}}%
\put(360,1660){\makebox(0,0)[r]{ 10000}}%
\put(360,1376){\makebox(0,0)[r]{ 1000}}%
\put(360,1092){\makebox(0,0)[r]{ 100}}%
\put(360,808){\makebox(0,0)[r]{ 10}}%
\put(360,524){\makebox(0,0)[r]{ 1}}%
\put(360,240){\makebox(0,0)[r]{ 0.1}}%
\end{picture}%
\endgroup
 
\caption{Star count predictions in the K band (2.15 $\mu$m) at
  l=0$^\circ$, for latitudes 10$^\circ$ to 90$^\circ$
from top to bottom (20$^\circ$, 45$^\circ$ and 90$^\circ$ with solid
  lines,  10$^\circ$, 30,$^\circ$ 60$^\circ$ with dashed 
lines). Data are from the DENIS point source catalogue in indicated  
directions.}
\end{center}
\end{figure*}

\begin{figure*}
\begin{center}
\begingroup%
  \makeatletter%
  \newcommand{\GNUPLOTspecial}{%
    \@sanitize\catcode`\%=14\relax\special}%
  \setlength{\unitlength}{0.1bp}%
\begin{picture}(4320,2592)(0,0)%
\special{psfile=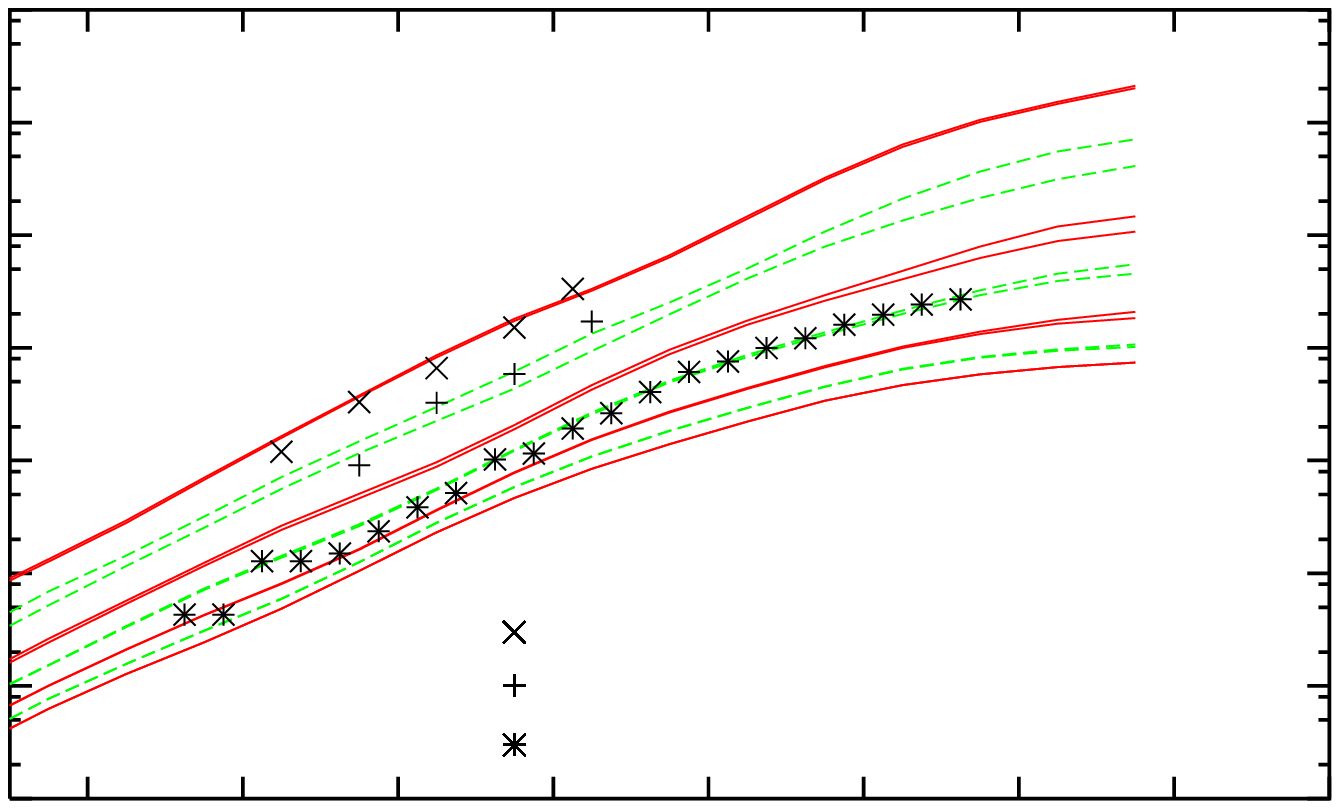 llx=0 lly=0 urx=432 ury=259 rwi=4320}
\fontsize{16}{\baselineskip}\selectfont
\put(1965,395){\makebox(0,0)[l]{Kummel and Wagner (96$^o$,30$^o$)}}%
\put(1965,719){\makebox(0,0)[l]{DENIS (270$^o$,0$^o$)}}%
\put(1965,565){\makebox(0,0)[l]{DENIS (270$^o$,-10$^o$)}}%
\put(3753,1466){\makebox(0,0)[l]{90$^o$}}%
\put(3753,1889){\makebox(0,0)[l]{20$^o$}}%
\put(3753,2299){\makebox(0,0)[l]{0$^o$}}%
\put(624,2299){\makebox(0,0)[l]{l=90$^o$ or 270$^o$ }}%
\put(2300,40){\makebox(0,0){K}}%
\put(4200,160){\makebox(0,0){ 22}}%
\put(3753,160){\makebox(0,0){ 20}}%
\put(3306,160){\makebox(0,0){ 18}}%
\put(2859,160){\makebox(0,0){ 16}}%
\put(2412,160){\makebox(0,0){ 14}}%
\put(1965,160){\makebox(0,0){ 12}}%
\put(1518,160){\makebox(0,0){ 10}}%
\put(1071,160){\makebox(0,0){ 8}}%
\put(624,160){\makebox(0,0){ 6}}%
\put(360,2512){\makebox(0,0)[r]{ 1e+06}}%
\put(360,2187){\makebox(0,0)[r]{ 100000}}%
\put(360,1863){\makebox(0,0)[r]{ 10000}}%
\put(360,1538){\makebox(0,0)[r]{ 1000}}%
\put(360,1214){\makebox(0,0)[r]{ 100}}%
\put(360,889){\makebox(0,0)[r]{ 10}}%
\put(360,565){\makebox(0,0)[r]{ 1}}%
\put(360,240){\makebox(0,0)[r]{ 0.1}}%
\end{picture}%
\endgroup
 
\caption{Star count predictions in the K band (2.15 $\mu$m) at
  l=90$^\circ$ or  270$^\circ$, for latitudes 0$^\circ$ to 90$^\circ$
from top to bottom (0$^\circ$, 20$^\circ$, 45$^\circ$ and 90$^\circ$
with solid  lines, 10$^\circ$, 30$^\circ$, 60$^\circ$ with dashed 
lines). For each latitude the bifurcation
is due to the warp, the highest part being for l=90, b$>$0 or l=270, b$<$0. 
Data are from the DENIS point source catalogue in indicated directions,
and from \protect\cite{Kummel2001} towards the North Ecliptic Pole.}
\end{center}
\end{figure*}

\begin{figure*}
\begin{center}
\begingroup%
  \makeatletter%
  \newcommand{\GNUPLOTspecial}{%
    \@sanitize\catcode`\%=14\relax\special}%
  \setlength{\unitlength}{0.1bp}%
\begin{picture}(4320,2592)(0,0)%
\special{psfile=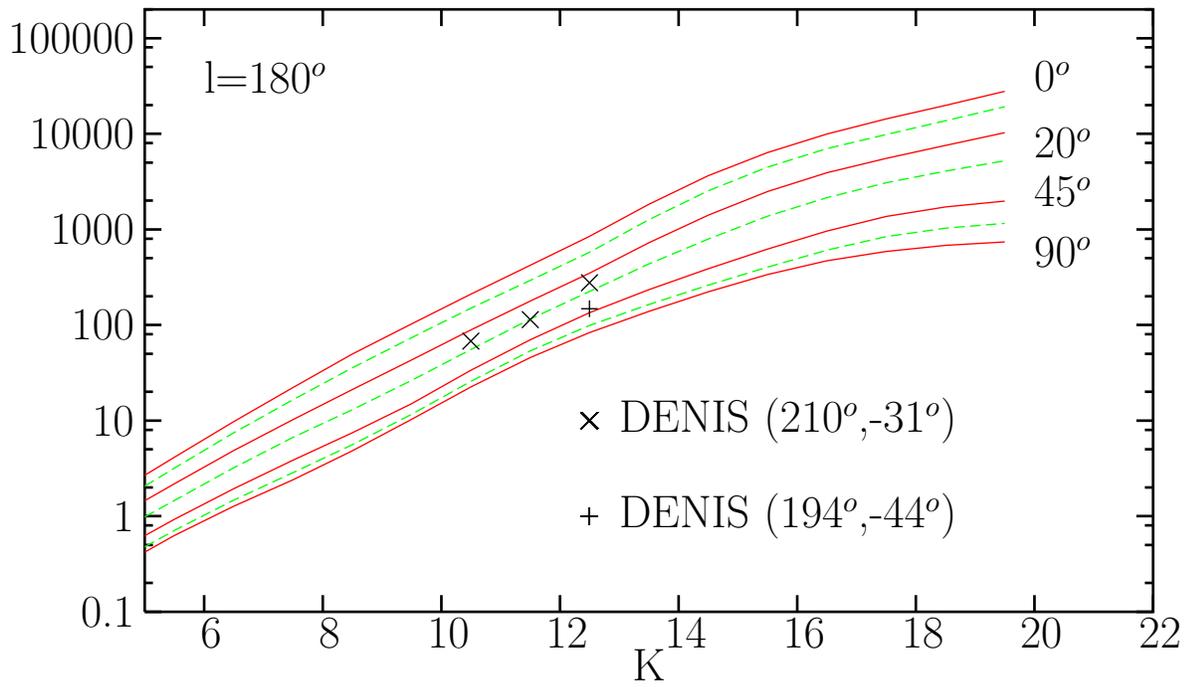 llx=0 lly=0 urx=432 ury=259 rwi=4320}
\fontsize{16}{\baselineskip}\selectfont
\put(2188,961){\makebox(0,0)[l]{DENIS (210$^o$,-31$^o$)}}%
\put(2188,601){\makebox(0,0)[l]{DENIS (194$^o$,-44$^o$)}}%
\put(3753,1597){\makebox(0,0)[l]{90$^o$}}%
\put(3753,1824){\makebox(0,0)[l]{45$^o$}}%
\put(3753,2008){\makebox(0,0)[l]{20$^o$}}%
\put(3753,2260){\makebox(0,0)[l]{0$^o$}}%
\put(624,2260){\makebox(0,0)[l]{l=180$^o$}}%
\put(2300,40){\makebox(0,0){K}}%
\put(4200,160){\makebox(0,0){ 22}}%
\put(3753,160){\makebox(0,0){ 20}}%
\put(3306,160){\makebox(0,0){ 18}}%
\put(2859,160){\makebox(0,0){ 16}}%
\put(2412,160){\makebox(0,0){ 14}}%
\put(1965,160){\makebox(0,0){ 12}}%
\put(1518,160){\makebox(0,0){ 10}}%
\put(1071,160){\makebox(0,0){ 8}}%
\put(624,160){\makebox(0,0){ 6}}%
\put(360,2403){\makebox(0,0)[r]{ 100000}}%
\put(360,2043){\makebox(0,0)[r]{ 10000}}%
\put(360,1682){\makebox(0,0)[r]{ 1000}}%
\put(360,1322){\makebox(0,0)[r]{ 100}}%
\put(360,961){\makebox(0,0)[r]{ 10}}%
\put(360,601){\makebox(0,0)[r]{ 1}}%
\put(360,240){\makebox(0,0)[r]{ 0.1}}%
\end{picture}%
\endgroup
 
\caption{Same as figure~8 but at l=180$^\circ$}
\end{center}
\end{figure*}

\begin{figure*}
\begin{center}
\begingroup%
  \makeatletter%
  \newcommand{\GNUPLOTspecial}{%
    \@sanitize\catcode`\%=14\relax\special}%
  \setlength{\unitlength}{0.1bp}%
\begin{picture}(4320,2592)(0,0)%
\special{psfile=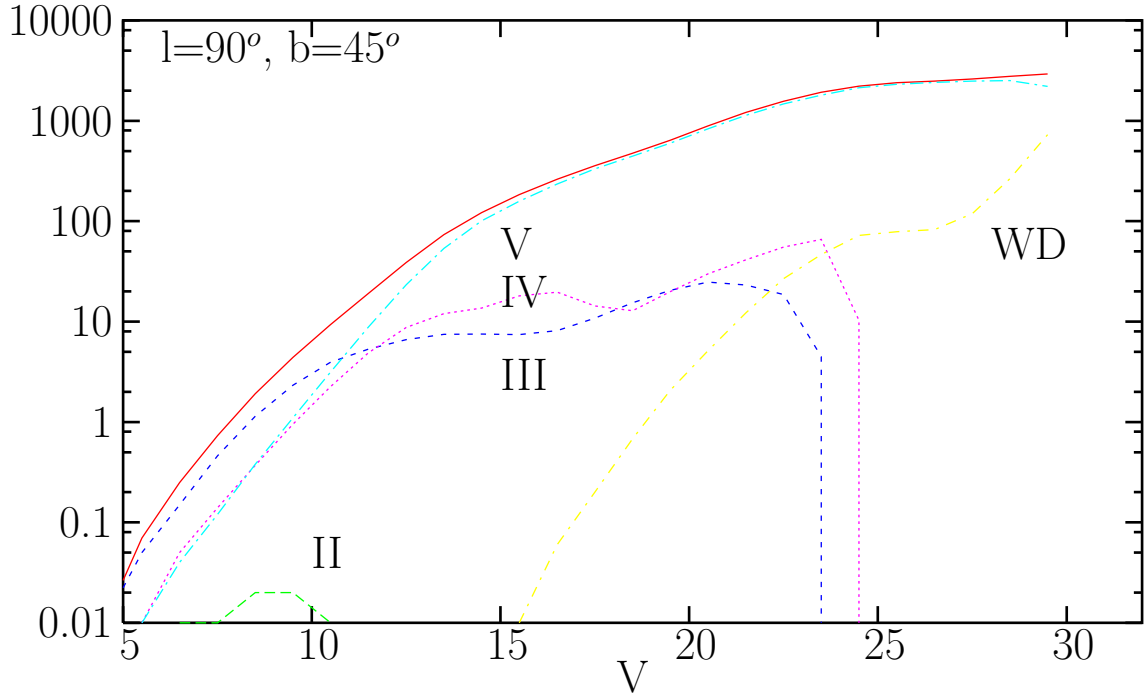 llx=0 lly=0 urx=432 ury=259 rwi=4320}
\fontsize{16}{\baselineskip}\selectfont
\put(3631,1671){\makebox(0,0)[l]{WD}}%
\put(1782,1671){\makebox(0,0)[l]{V}}%
\put(1782,1490){\makebox(0,0)[l]{IV}}%
\put(1782,1178){\makebox(0,0)[l]{III}}%
\put(1071,505){\makebox(0,0)[l]{II}}%
\put(502,2398){\makebox(0,0)[l]{l=90$^o$, b=45$^o$}}%
\put(2280,40){\makebox(0,0){V}}%
\put(3916,160){\makebox(0,0){ 30}}%
\put(3204,160){\makebox(0,0){ 25}}%
\put(2493,160){\makebox(0,0){ 20}}%
\put(1782,160){\makebox(0,0){ 15}}%
\put(1071,160){\makebox(0,0){ 10}}%
\put(360,160){\makebox(0,0){ 5}}%
\put(320,2512){\makebox(0,0)[r]{ 10000}}%
\put(320,2133){\makebox(0,0)[r]{ 1000}}%
\put(320,1755){\makebox(0,0)[r]{ 100}}%
\put(320,1376){\makebox(0,0)[r]{ 10}}%
\put(320,997){\makebox(0,0)[r]{ 1}}%
\put(320,619){\makebox(0,0)[r]{ 0.1}}%
\put(320,240){\makebox(0,0)[r]{ 0.01}}%
\end{picture}%
\endgroup
 
\caption{Star count predictions (stars per magnitude and per 
square degree) in the V band at l=90$^\circ$ and b=45$^\circ$ and
contributions  of the different luminosity classes. All: solid line,
class II:  long dashed, class III: short dashed, class IV: dotted,
class V:  dashed-dotted upper curve, white dwarfs: dashed-dotted lower
curve.  The giants dominate at V$<$11, main sequence stars at V$>14$
in this direction. The rise of the white
dwarf curve at V$>27$ is due to the dark halo white dwarf population
normalized here to 2\% of the dark halo.}
\end{center}
\end{figure*}

\begin{figure*}
\begin{center}
\hbox{\hspace{0cm}
\begingroup%
  \makeatletter%
  \newcommand{\GNUPLOTspecial}{%
    \@sanitize\catcode`\%=14\relax\special}%
  \setlength{\unitlength}{0.1bp}%
\begin{picture}(2520,1512)(0,0)%
\special{psfile=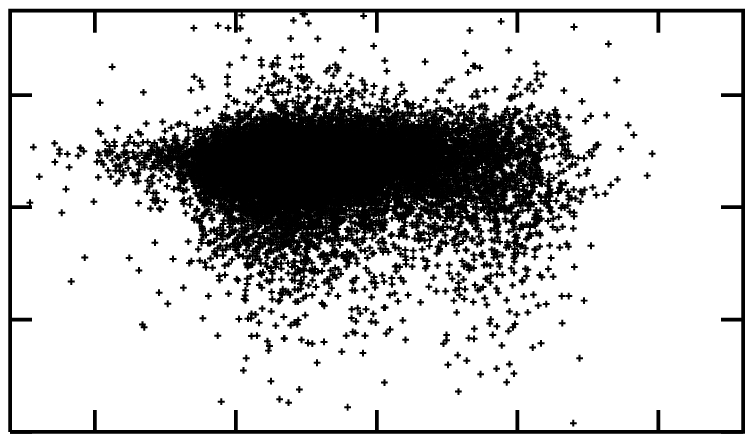 llx=0 lly=0 urx=252 ury=151 rwi=2520}
\fontsize{15}{\baselineskip}\selectfont
\put(622,306){\makebox(0,0)[l]{a}}%
\put(1352,38){\makebox(0,0){B-V}}%
\put(74,831){%
\special{ps: gsave currentpoint currentpoint translate
270 rotate neg exch neg exch translate}%
\makebox(0,0)[b]{\shortstack{$\mu_l$}}%
\special{ps: currentpoint grestore moveto}%
}%
\put(2164,150){\makebox(0,0){ 2}}%
\put(1758,150){\makebox(0,0){ 1.5}}%
\put(1353,150){\makebox(0,0){ 1}}%
\put(947,150){\makebox(0,0){ 0.5}}%
\put(541,150){\makebox(0,0){ 0}}%
\put(260,1195){\makebox(0,0)[r]{ 2}}%
\put(260,872){\makebox(0,0)[r]{-2}}%
\put(260,548){\makebox(0,0)[r]{-6}}%
\put(260,225){\makebox(0,0)[r]{-10}}%
\end{picture}%
\endgroup
 
\hspace{0cm}
\begingroup%
  \makeatletter%
  \newcommand{\GNUPLOTspecial}{%
    \@sanitize\catcode`\%=14\relax\special}%
  \setlength{\unitlength}{0.1bp}%
\begin{picture}(2520,1512)(0,0)%
\special{psfile=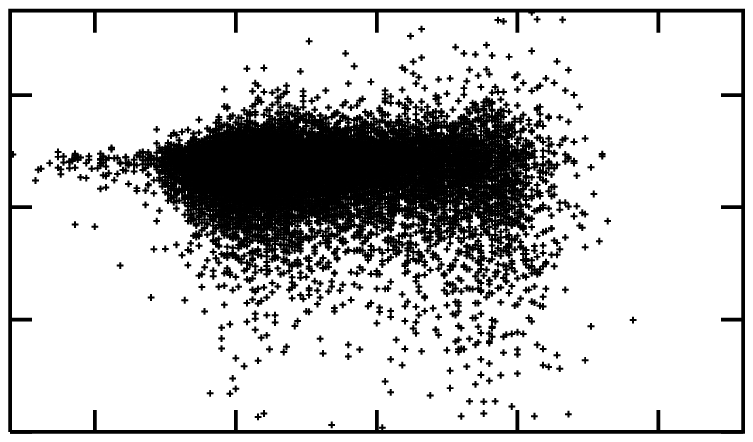 llx=0 lly=0 urx=252 ury=151 rwi=2520}
\fontsize{15}{\baselineskip}\selectfont
\put(622,306){\makebox(0,0)[l]{b}}%
\put(1352,38){\makebox(0,0){B-V}}%
\put(74,831){%
\special{ps: gsave currentpoint currentpoint translate
270 rotate neg exch neg exch translate}%
\makebox(0,0)[b]{\shortstack{$\mu_l$}}%
\special{ps: currentpoint grestore moveto}%
}%
\put(2164,150){\makebox(0,0){ 2}}%
\put(1758,150){\makebox(0,0){ 1.5}}%
\put(1353,150){\makebox(0,0){ 1}}%
\put(947,150){\makebox(0,0){ 0.5}}%
\put(541,150){\makebox(0,0){ 0}}%
\put(260,1195){\makebox(0,0)[r]{ 2}}%
\put(260,872){\makebox(0,0)[r]{-2}}%
\put(260,548){\makebox(0,0)[r]{-6}}%
\put(260,225){\makebox(0,0)[r]{-10}}%
\end{picture}%
\endgroup
 
\hfill
}
\caption{B-V versus proper motion $\mu_l$ diagrams towards
  l=44$^\circ$,  b=474$^\circ$, for
stars with $12<V<18$. a) Observations from \protect\cite{Ojha1996}. b)
  Simulation.}
\end{center}
\end{figure*}

\begin{acknowledgements}
{ We are grateful to G. Bruzual, P. Bergeron, E. Garcia-Berro and G. 
Chabrier for giving us their models in tabular form, and to 
H. Jahrei\ss~for 
furnishing an up-to-date version 
of the local luminosity function prior to publication. We also thank 
the referee for pertinent comments and suggestions.
SP has benefited from an allowance from the Région de Franche-Comté.}
\end{acknowledgements}

\end{document}